\documentclass[aps,pra,twocolumn,showpacs,superscriptaddress,nofootinbib]{revtex4-2}
\usepackage{amsmath,amssymb,amsfonts}
\allowdisplaybreaks[4]
\usepackage{graphicx}
\usepackage{dcolumn}
\usepackage{epstopdf}
\usepackage[T1]{fontenc}
\usepackage{newtxtext}
\usepackage{subfigure}
\usepackage[percent]{overpic}
\usepackage{float}
\usepackage{bm}
\usepackage{algorithm}
\usepackage{algpseudocode}
\usepackage{booktabs}
\usepackage{float}

\allowdisplaybreaks

\usepackage[unicode]{hyperref}
\hypersetup{
	unicode=true,          % non-Latin characters in Acrobat bookmarks
	a4paper=true,
	plainpages=false,
	pdftitle={Title of PDF},    % title
	pdfauthor={Author of PDF},     % author
	pdfsubject={Subject of PDF},   % subject of the document
	colorlinks=true,       % false: boxed links; true: colored links
	linkcolor=blue,          % color of internal links
	citecolor=blue,        % color of links to bibliography
	filecolor=blue,      % color of file links
	urlcolor=blue           % color of external links
}
\newcommand{\reffig}[1]{Fig.~\ref{#1}}

\newcommand{\reftable}[1]{Table~\ref{#1}}

\DeclareMathAlphabet{\mathcal}{OMS}{cmsy}{m,b}{n,it}

\begin{document}

	\title{Thermal effects on density-modulated phases in a dipolar Bose--Einstein condensate} 
	
	\author{Changjian Yu}
	\author{Jinbin Li}
	\author{Kui-Tian Xi}
	\email[Corresponding author: ]{xiphys@nuaa.edu.cn}
	\affiliation{College of Physics, Nanjing University of Aeronautics and Astronautics, Nanjing, 211106, China}
	\affiliation{Key Laboratory of Aerospace Information Materials and Physics (Nanjing University of Aeronautics and Astronautics), MIIT, Nanjing, 211106, China}
	
	\date{\today}

	\begin{abstract}
	Density-modulated phases in dipolar Bose--Einstein condensates arise from the competition between contact repulsion, long-range dipole--dipole interactions, and beyond-mean-field fluctuations. We examine finite-temperature stationary states of a harmonically trapped dipolar Bose gas using a temperature-dependent extended Gross--Pitaevskii equation within the local-density approximation. In this framework, the thermal correction shifts some structural boundaries toward larger scattering lengths and changes the parameter region in which connected density-modulated states are obtained. The density contrast changes sharply at $T = 0$, whereas at finite temperature it varies more gradually, consistent with a smooth crossover-like evolution in the finite trapped system. We also use Leggett's bound to estimate a geometric upper limit on the possible superfluid response for representative density profiles. These results suggest that finite-temperature corrections can reshape pattern formation in dipolar quantum gases.
	\end{abstract}
	
	%\pacs{}
	%\keywords{}
	
	\maketitle
	
	%%%%%%%%%%%%%%%%%%%%%%%%%%%%%%%%%%%%%%%%%%%%%%%%%%%%%%%%%%%%%%%%%%%%%%

	\section{Introduction}
	
	Competing interactions often produce spatially modulated structures, including droplets, honeycomb patterns, and labyrinthine networks. Related morphologies appear in systems ranging from quantum fluids to classical ferrofluids, and analogous ``nuclear pasta'' structures have been discussed in neutron-star crusts \cite{Lahaye2009,YKora2019,FBottcher2021,Stavropoulos1998,Dalfovo2001,Ravenhall1983,Chamel2008}. Dipolar Bose--Einstein condensates (dBECs) provide a clean and controllable setting in which to study such pattern formation, because the contact interaction, the long-range anisotropic dipole--dipole interaction, and the trapping geometry can be tuned with high precision \cite{Ferrier-Barbut2016,Schmitt2016,Kadau2016}. In these gases, the same ingredients also underlie roton softening \cite{Tanzi2019,Guo2019,Natale2019,Bottcher2019,Chomaz2018}, superfluid--supersolid transitions \cite{Tanzi2019,Chomaz2019,Ilzhofer2021}, and two-dimensional pattern formation \cite{Hertkorn2021,Zhang2021,Tanzi2019prl, Norcia2021}.
	
	Beyond-mean-field quantum fluctuations play an essential role in this physics. The Lee--Huang--Yang (LHY) correction provides a repulsive contribution that can stabilize dipolar gases against mean-field collapse \cite{Lima2011,Lima2012}, enabling quantum droplets and self-organized density textures with close analogs in classical ferrofluids \cite{Seul1995,Rosensweig1997,Andelman2009,Bourgine2011}. When density modulation is accompanied by phase coherence and a finite superfluid response, such states may realize supersolidity. Experiments with strongly dipolar gases have shown that modulated and supersolid-like states can arise intrinsically from dipolar interactions supplemented by quantum fluctuations \cite{Kadau2016,Tanzi2019,Guo2019,Natale2019,Bottcher2019,Chomaz2019}. These studies have motivated detailed zero-temperature structural maps including uniform condensates, droplet arrays, and connected density-modulated states \cite{Hertkorn2021,Zhang2024}.
	
	Finite temperature adds another level of control, but it also introduces additional theoretical limitations. Thermal fluctuations are often associated with decoherence and depletion. Recent work, however, indicates that they can also modify the stability of modulated states in dipolar gases \cite{Griffin1996,Ronen2007,Bisset2012,Ticknor2012,Pawlowski2013,Aybar2019,Ozturk2020,Sohmen2021,SanchezBaena2023,SanchezBaena2024,He2025,Bombin2025}. In particular, temperature-dependent behavior has been reported near the boundary between homogeneous and density-modulated regimes \cite{SanchezBaena2023}. This motivates the more focused question addressed here: within a finite-temperature mean-field description, how does the thermal correction modify stationary density patterns and their structural boundaries in a trapped dipolar condensate?
	
	We address this question for a single-component dipolar BEC in a cylindrically symmetric harmonic trap. The calculation uses a temperature-dependent extended Gross--Pitaevskii equation (TeGPE) within the local-density approximation (LDA), with both LHY and thermal fluctuation corrections. The resulting structural map indicates that the thermal term shifts some boundaries toward larger scattering lengths and changes the region in which connected density-modulated stationary states are obtained. We also track the density contrast as a function of scattering length and atom number. At $T=0$ the contrast changes sharply, whereas at finite temperature it varies more smoothly; for the finite trapped system considered here, we describe this behavior as a smooth crossover-like evolution unless further thermodynamic evidence is supplied. Finally, we use Leggett's bound only as a density-based upper bound on the possible superfluid response, not as a standalone proof of supersolidity.
	
	The paper is organized as follows. Sec.~\ref{SEC:Formalism} introduces the finite-temperature extended Gross--Pitaevskii framework. Secs.~\ref{SEC:PHASE DIAGRAM} and \ref{SEC:PHASE TRANSITION} present the structural map and the density-contrast analysis. Sec.~\ref{SEC:Leggett-bound estimate} gives the Leggett-bound estimate. We summarize the main results and limitations in Sec.~\ref{SEC:Conclusion}.
	
	\section{Formalism}\label{SEC:Formalism}
	
	We consider a dipolar BEC in a cylindrically symmetric harmonic trap. The stationary solutions are obtained from the temperature-dependent extended Gross--Pitaevskii equation (TeGPE) \cite{Aybar2019,Ozturk2020}. In its time-dependent form,
	\begin{widetext}
		\begin{equation}
			i \hbar \frac{\partial}{\partial t} \Psi(\bm{r}, t) = \bigg[ -\frac{\hbar^{2}}{2 m} \nabla^{2} + V(\bm{r}) + g|\Psi(\bm{r}, t)|^2
			+ \int U\left(\bm{r}-\bm{r^{\prime}}\right) |\Psi(\bm{r^{\prime}}, t)|^{2} d^3 r^{\prime}
			+ H_{\text{qu}}(\bm{r}) + H_{\text{th}}(\bm{r}) \bigg]  \Psi(\bm{r}, t),
			\label{eqn:TeGPE}
		\end{equation}
	\end{widetext}
	where $m$ is the atomic mass, $a_s$ is the $s$-wave scattering length, and $g=4\pi\hbar^2a_s/m$ is the contact coupling. The dipole--dipole interaction is $U(\bm r-\bm r')=\mu_0\mu^2(1-3\cos^2\theta)/(4\pi|\bm r-\bm r'|^3)$, with the dipoles polarized along the $z$ direction and $\theta$ the angle between $\bm r-\bm r'$ and the polarization axis. The trap is $V(\bm r)=m(\omega_x^2x^2+\omega_y^2y^2+\omega_z^2z^2)/2$, with $(\omega_x,\omega_y,\omega_z)=2\pi\times(125,125,250)$ Hz. The order parameter is normalized to the condensate atom number, $N=\int |\Psi(\bm r,t)|^2d^3r$. The terms $H_{\text{qu}}$ and $H_{\text{th}}$ denote, respectively, the quantum and thermal fluctuation corrections used in Refs.~\cite{Lima2012,SanchezBaena2023,Aybar2019,Ozturk2020}.
	
	The quantum-fluctuation term is taken in the usual local-density form of the LHY correction \cite{Lee1051957,Lee1061957,Lima2012}. This approximation has been benchmarked against quantum Monte Carlo calculations and has been widely used to describe dipolar droplets and supersolid-candidate states \cite{Saito2016,Tanzi2019,Guo2019,Natale2019,Bottcher2019,Chomaz2019}. We use
	\begin{align}
		H_{\text{qu}}(\bm{r}) &= \frac{32}{3} g \sqrt{\frac{a_{s}^{3}}{\pi}} \left( 1 + \frac{3}{2} \varepsilon_{dd}^{2} \right) |\Psi(\bm{r}, t)|^{3} ,
		\label{eqn:quantum fluctuation}
	\end{align}
	where $\varepsilon_{dd}=a_{dd}/a_s$, and $a_{dd}$ is the dipolar length.

	The thermal correction is written as \cite{Aybar2019}
	\begin{align}
		H_{\text{th}}(\bm{r}) &= \theta T^2\frac{1}{|\Psi(\bm{r})|},
		\label{eqn:thermal fluctuation}
	\end{align}
	where $\theta=(32/3)g\sqrt{a_s^3/\pi}\,(k_B^2/g^2)\,\mathcal{S}(\epsilon_{dd})$, with 
	\begin{align}
		\mathcal{S}(\epsilon_{dd}) = & -0.01029\epsilon_{dd}^4 + 0.02963\epsilon_{dd}^3 \nonumber\\ 
		& -0.05422\epsilon_{dd}^2 + 0.009302\epsilon_{dd} + 0.1698.
	\end{align}
	The derivation of Eq.~\eqref{eqn:thermal fluctuation} uses a local-density treatment of the thermal Bogoliubov contribution and requires an infrared cutoff when $\epsilon_{dd}>1$ \cite{Aybar2019}. In a finite trapped cloud, wavelengths much larger than the local high-density region are not supported as local excitations. We therefore use the cutoff $k_c\simeq\pi/(2\xi)$, with $\xi=\hbar/\sqrt{2mgn}$ the local healing length. This prescription follows Ref.~\cite{Aybar2019}. It should be viewed as part of the TeGPE--LDA prescription rather than as a unique microscopic treatment of all low-energy modes; quantitative boundary locations may therefore depend on improvements such as anisotropic cutoffs or nonlocal finite-temperature theories. The $1/|\Psi|$ form should not be read literally in the extreme low-density tail, where the LDA itself is no longer controlled. We therefore base the structural classification on the dense part of the cloud. The threshold used later in the density-contrast analysis is a post-processing definition of $C$, not an additional modification of the TeGPE evolution. A derivation of the fitted form is summarized in Appendix~\ref{appderive}.
	
	Stationary states are obtained by imaginary-time propagation. After each time step the order parameter is renormalized to the prescribed condensate atom number. We monitor convergence using the total energy and chemical potential. The numerical calculation uses a Fourier pseudospectral discretization on a Cartesian grid: local terms are evaluated in real space, while the nonlocal dipolar convolution is computed in momentum space using fast Fourier transforms.
	
	\section{Numerical Results}\label{SEC:NUMERICAL RESULTS}
	\subsection{Structural map}\label{SEC:PHASE DIAGRAM}
	
	We first construct a finite-temperature structural map and compare it with the zero-temperature results of Ref.~\cite{Hertkorn2021}. The map should be understood as a classification of stationary density morphologies within the present TeGPE--LDA model, rather than as a complete thermodynamic phase diagram. We consider a strongly dipolar gas of $^{162}$Dy atoms with dipolar length $a_{dd}=130a_0$ in an oblate harmonic trap, $(\omega_x,\omega_y,\omega_z)=2\pi\times(125,125,250)$ Hz. This geometry favors quasi-two-dimensional density modulation.

\begin{figure}[t]
		\centering
		\subfigure{
			\includegraphics[width=0.44\textwidth]{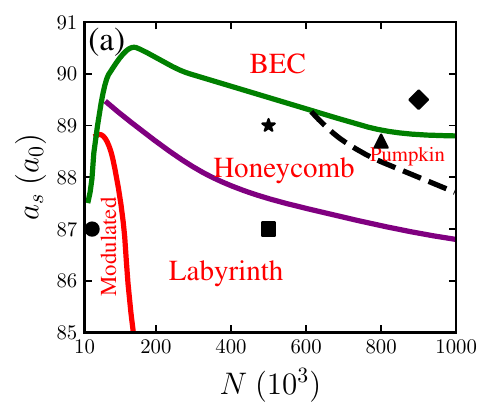}
		}\\[0em]
		\subfigure{
			\includegraphics[width=0.4\textwidth]{phase_morphologies100nk.pdf}
		}\\[0em]
		\subfigure{
			\includegraphics[width=0.44\textwidth]{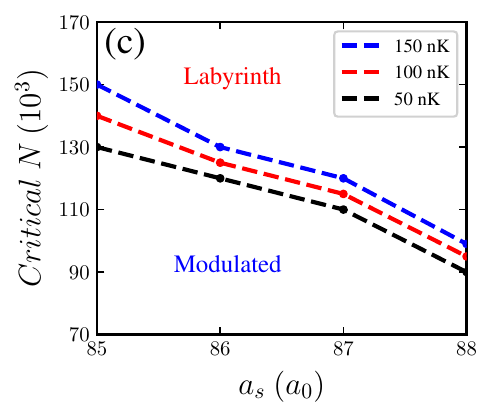}
		}
		\caption{Finite-temperature structural map at $T=100$ nK for a $^{162}$Dy BEC in a harmonic trap with $(\omega_x,\omega_y,\omega_z)=2\pi\times(125,125,250)$ Hz. (a) Structural boundaries in the $(N,a_s)$ plane. (b) Representative column-density profiles $\rho_{2\text{D}}(x,y)=\int |\Psi(\bm r)|^2dz$, illustrating the uniform BEC, connected density-modulated, honeycomb, labyrinthine, and pumpkin-like morphologies. (c) Temperature-dependent critical condensate number for the boundary between connected density-modulated and labyrinthine states at $T=50$, $100$, and $150$ nK.}
		\label{phase_diagram}
	\end{figure}

	At $T=100$ nK we scan the $s$-wave scattering length $a_s$ and condensate number $N$, as summarized in \reffig{phase_diagram}(a). Representative column-density profiles, $\rho_{2\text{D}}(x,y)=\int |\Psi(\bm r)|^2dz$, are shown in \reffig{phase_diagram}(b). The profiles include a nearly uniform BEC, connected density-modulated states, honeycomb networks, labyrinthine patterns, and a pumpkin-like morphology. For the representative points shown, the local dimensionless temperature $t=k_BT/(gn_0)$, evaluated at the peak density $n_0$, lies within $0<t<10$, the range used in deriving the fitted thermal correction \cite{Aybar2019}. Each stationary state is obtained by imaginary-time propagation from a randomized Perlin-noise seed. The labels in Fig.~\ref{phase_diagram} are assigned from the real-space morphology, the connectivity of density maxima, and the density contrast defined below. They should therefore be understood as structural classifications of stationary solutions rather than sharp thermodynamic phase boundaries.
	
	As $a_s$ is lowered, the initially smooth density distribution develops finite-wavelength modulations. Depending on $(a_s,N)$, the stationary profiles range from droplet-like arrays to connected honeycomb and labyrinthine networks. Close to the onset of modulation and at smaller $N$, we also find ring-like and blood-cell-like profiles, consistent with earlier work on trapped dipolar gases \cite{Schmidt2021,Kawaguchi2012,Ronen2006,Eberlein2005,RonenL2007,Dutta2007,Wilson2008}. The honeycomb morphology is characterized by a connected network with approximate sixfold symmetry, while the labyrinthine morphology forms elongated, stripe-like domains with only short-range orientational order. The pumpkin-like state appears in an intermediate region and is identified here as a distinct structural morphology. These labels are therefore used as density-pattern classifications within the present finite trapped system.
	
	Relative to the zero-temperature structural map of Ref.~\cite{Hertkorn2021}, the finite-temperature correction shifts some of the relevant boundaries toward larger scattering lengths. For example, at $N=10^4$ the boundary between the density-modulated and BEC-like profiles occurs near $a_s\simeq87.5a_0$ at $T=100$ nK, and the rightmost extent of the modulated region reaches about $88.87a_0$. The upper boundary of the honeycomb region also moves to larger $a_s$, with a maximum near $90.5a_0$. Figure~\ref{phase_diagram}(c) shows a complementary trend for the boundary between connected density-modulated and labyrinthine states: the corresponding critical condensate number increases with temperature. Thus, finite temperature does not shift all boundaries in the same direction in the $(N,a_s)$ plane; instead, it reshapes the structural map in a state-dependent manner.
	
	This behavior reflects the way the thermal correction modifies the local equation of state. The LHY term remains the main repulsive beyond-mean-field contribution at high density, while the finite-temperature term changes the balance among contact, dipolar, and fluctuation energies. In the stationary solutions considered here, this modified balance changes where connected density-modulated profiles are obtained within the scanned parameter range. A more quantitative separation of energetic and entropic effects would require a finite-temperature treatment beyond the local stationary model used here.
	
	\begin{figure}[t]
		\centering
		\subfigure{
			\includegraphics[width=0.42\textwidth]{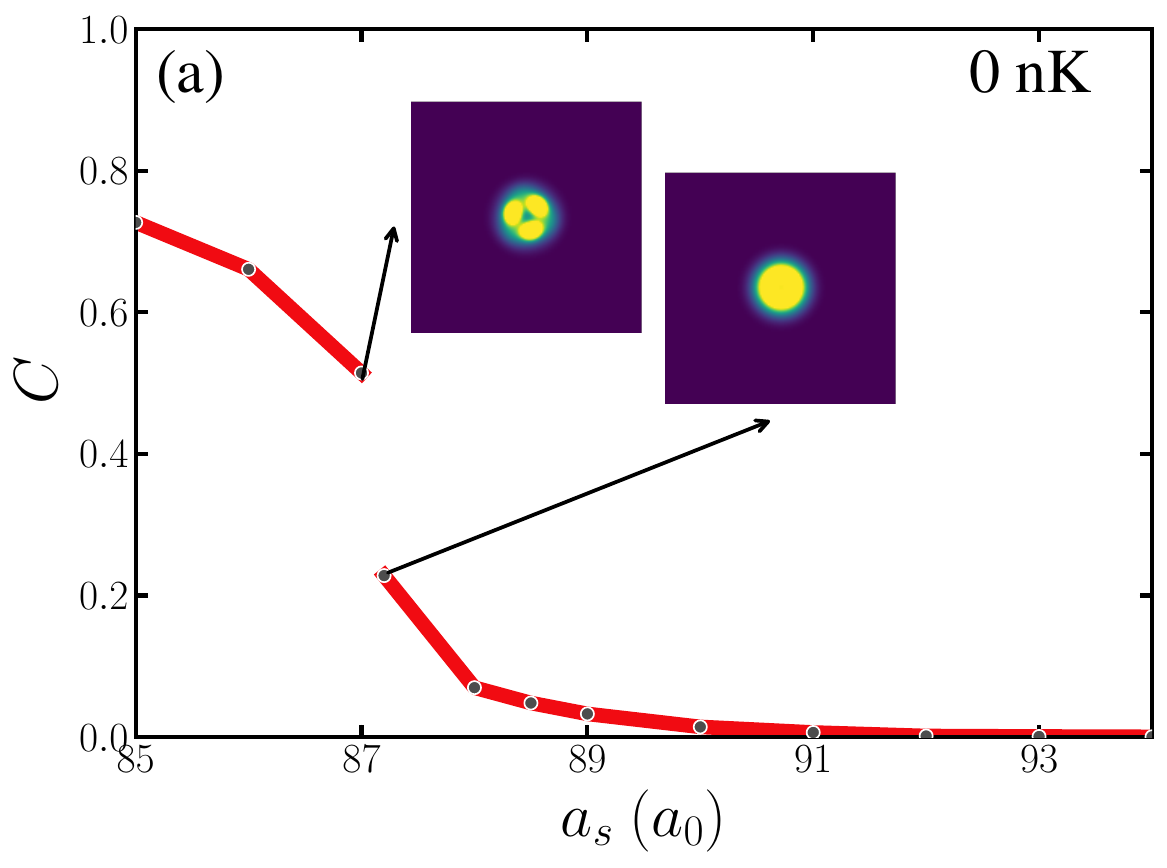}
		}\\[0em]
		\subfigure{
			\includegraphics[width=0.42\textwidth]{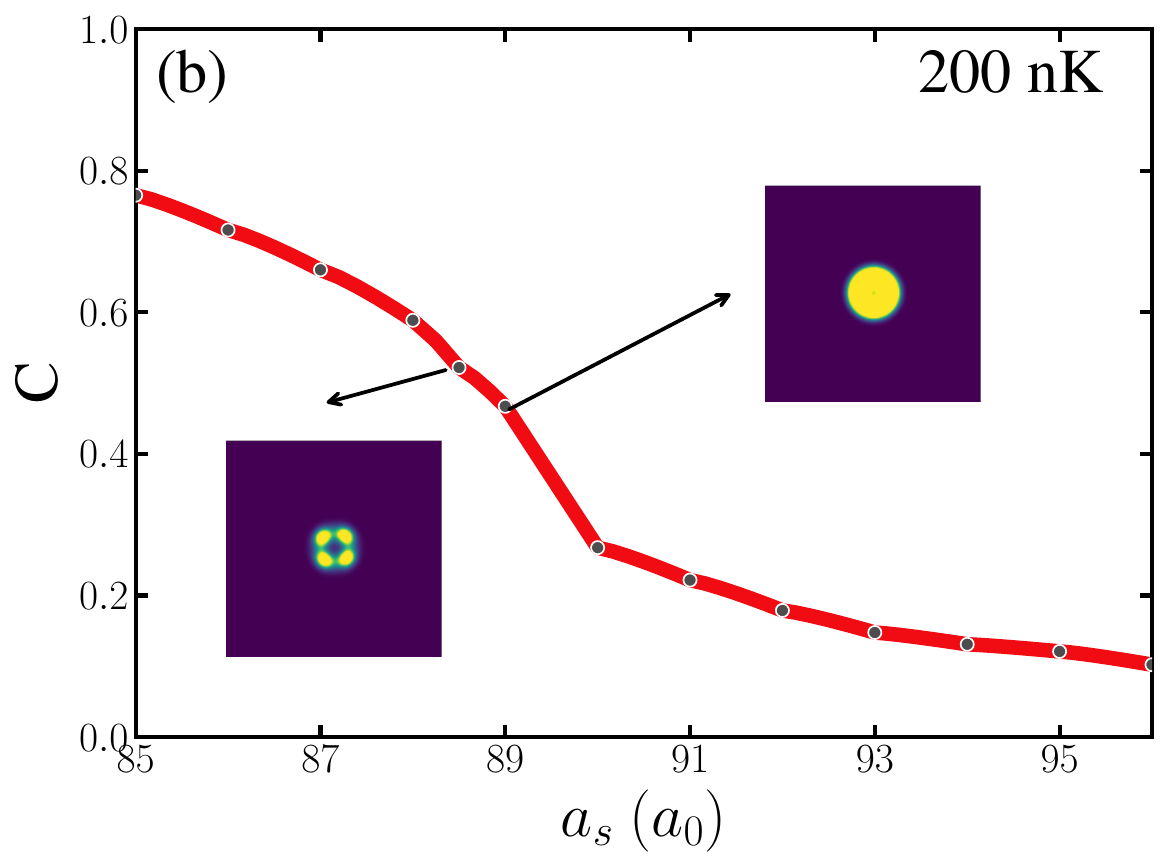}
		}
		\caption{Effect of temperature on the density-modulated to BEC-like crossover. (a),(b) Density contrast $C$ versus scattering length $a_s$ at $T=0$ and $T=200$ nK, respectively. Insets show representative column densities near the loss of visible modulation.}
		\label{C_vs_as}
	\end{figure}
	
	\subsection{Density-contrast analysis}\label{SEC:PHASE TRANSITION}

	To quantify the degree of modulation, we use the density contrast
	\begin{equation}
		C = \frac{\rho_{\text{max}} - \rho_{\text{min}}}{\rho_{\text{max}} + \rho_{\text{min}}},
		\label{eqn:density contrast}
	\end{equation}
	where $\rho_{\text{max}}$ and $\rho_{\text{min}}$ are extracted from the dimensionless normalized column density $\rho_{2\text{D}}=\int |\Psi(\bm r)|^2dz$. To avoid spurious contributions from the dilute numerical boundary, $\rho_{\text{min}}$ is taken only from the region satisfying $\rho_{2\text{D}}(x,y)>0.025$. This fixed threshold is used only in the post-processing evaluation of $C$; it is not introduced in the TeGPE evolution and is not used as an additional criterion for assigning the structural labels. A large value of $C$ signifies a pronounced density modulation, whereas $C \simeq 0$ indicates that the condensate is essentially uniform. Figure~\ref{C_vs_as} shows $C$ as a function of $a_s$ at fixed $N=2\times10^4$ for $T=0$ and $T=200$ nK. At $T=0$, $C$ stays close to unity until $a_s\simeq87a_0$ and then drops sharply between $87a_0$ and $87.2a_0$. At $T=200$ nK the decrease of $C$ occurs at larger scattering lengths, roughly $88.5$--$90a_0$, and is noticeably smoother. We therefore describe the zero-temperature curve as showing a sharp, discontinuous-like change, while the finite-temperature curve is consistent with a smooth crossover-like evolution in the finite trapped system. The $T=200$ nK cut is used to illustrate this trend in the stationary solutions rather than to define a high-precision thermodynamic boundary.
	
\begin{figure}[t]
		\centering
		\subfigure{
			\includegraphics[width=0.41\textwidth]{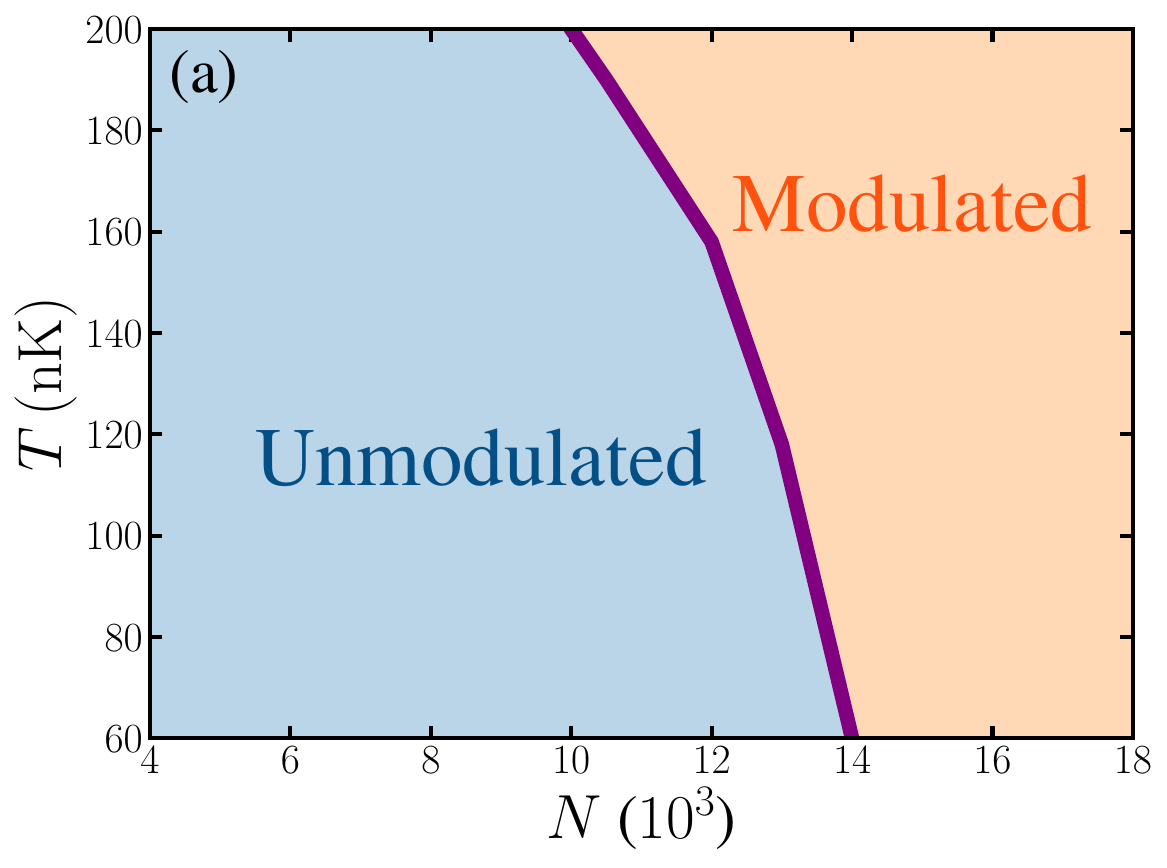}
		}\\[0em]
		\subfigure{
			\includegraphics[width=0.4\textwidth]{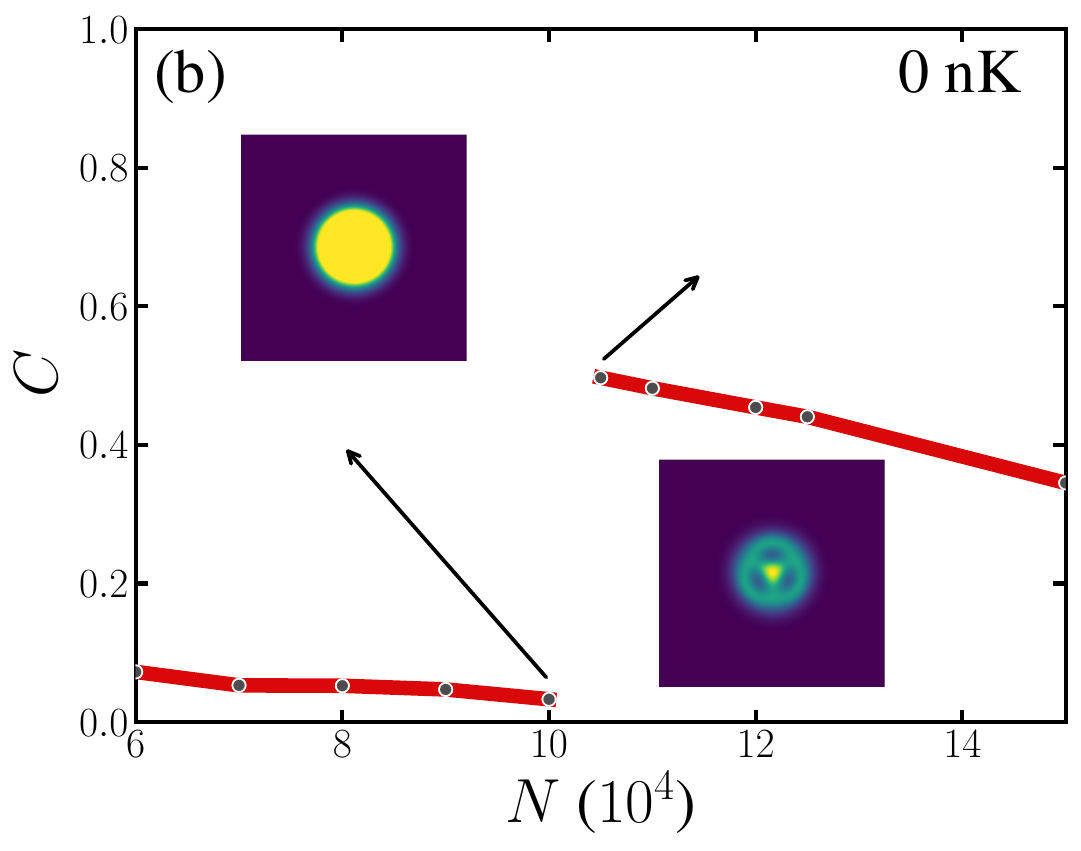}
		}\\[0em]
		\subfigure{
			\includegraphics[width=0.4\textwidth]{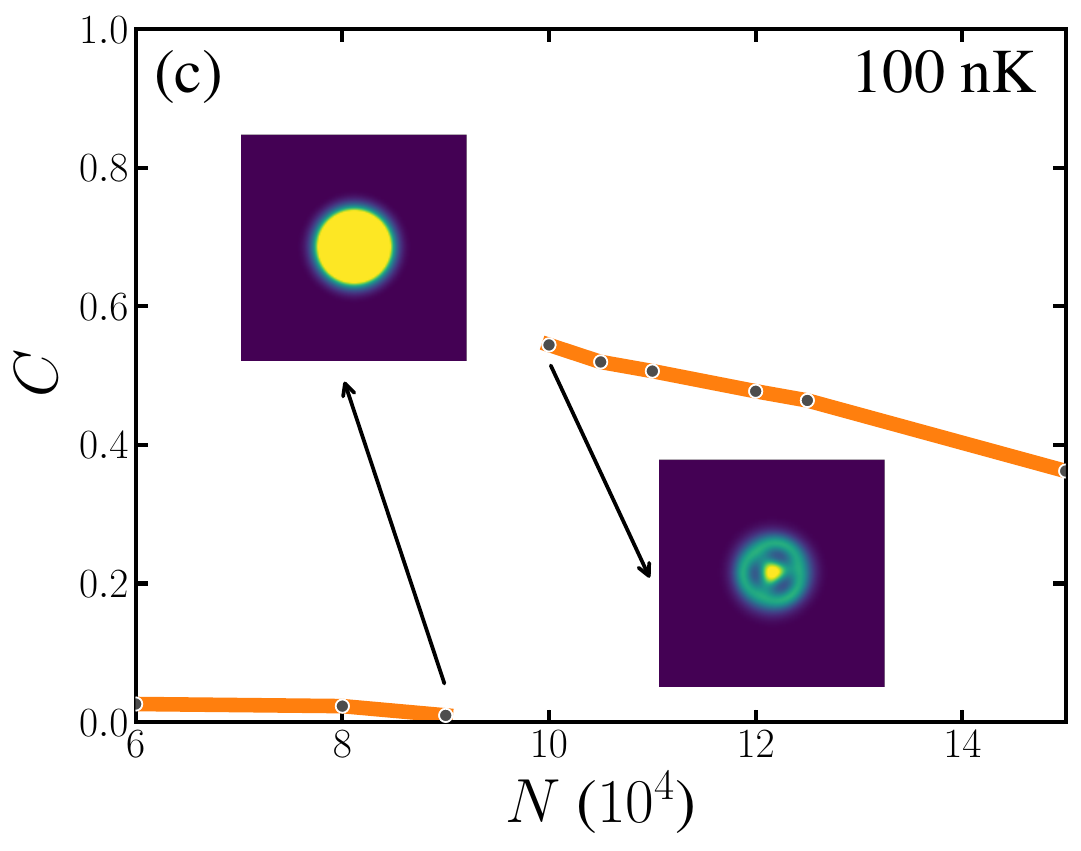}
		}
		\caption{Effect of temperature on the onset of density modulation. (a) Boundary in the temperature--condensate-number plane at fixed $a_s=85a_0$. (b),(c) Density contrast $C$ versus condensate number $N$ at $T=0$ and $T=100$ nK, respectively, for fixed $a_s=90a_0$. Insets show representative column densities.}
		\label{C_vs_N}
	\end{figure}

	Figure~\ref{C_vs_N} gives a complementary view at fixed scattering length. In \reffig{C_vs_N}(a), the boundary between unmodulated and modulated profiles in the $(T,N)$ plane is shown for $a_s=85a_0$. The critical condensate number decreases as temperature is raised. The cuts in \reffig{C_vs_N}(b) and \reffig{C_vs_N}(c), taken at $a_s=90a_0$, show that the onset of modulation moves from $N\simeq1.05\times10^5$ at $T=0$ to about $10^5$ at $T=100$ nK.
	
	Taken together, the two cuts show that the thermal term can lower the condensate number required for the onset of modulation, while other structural boundaries, such as the connected-modulated to labyrinthine boundary in \reffig{phase_diagram}(c), move in the opposite direction in $N$. This reinforces the view that finite temperature reshapes the stationary density landscape rather than simply stabilizing or destabilizing all modulated states uniformly.

	\subsection{Leggett-bound estimate}\label{SEC:Leggett-bound estimate}
	
	We finally estimate the possible superfluid response using Leggett's upper bound \cite{Leggett1970}. This quantity is a geometric bound determined by the density profile; it is a useful diagnostic of transport constraints, but it does not by itself establish supersolidity. For flow along $x$, the bound is written in terms of the one-dimensional density $n_{\rm 1D}(x)=\int n(\bm r)\,dy\,dz$ \cite{Leggett1970,Leggett1998}:
	\begin{equation}
		f_S \leqslant \left( \left\langle n_{\rm 1D}(x) \right\rangle \left\langle \frac{1}{n_{\rm 1D}(x)} \right\rangle \right)^{-1}.
		\label{eq:leggett_bound}
	\end{equation}
	
	\begin{figure}[t]
		\centering
		\includegraphics[width=0.43\textwidth]{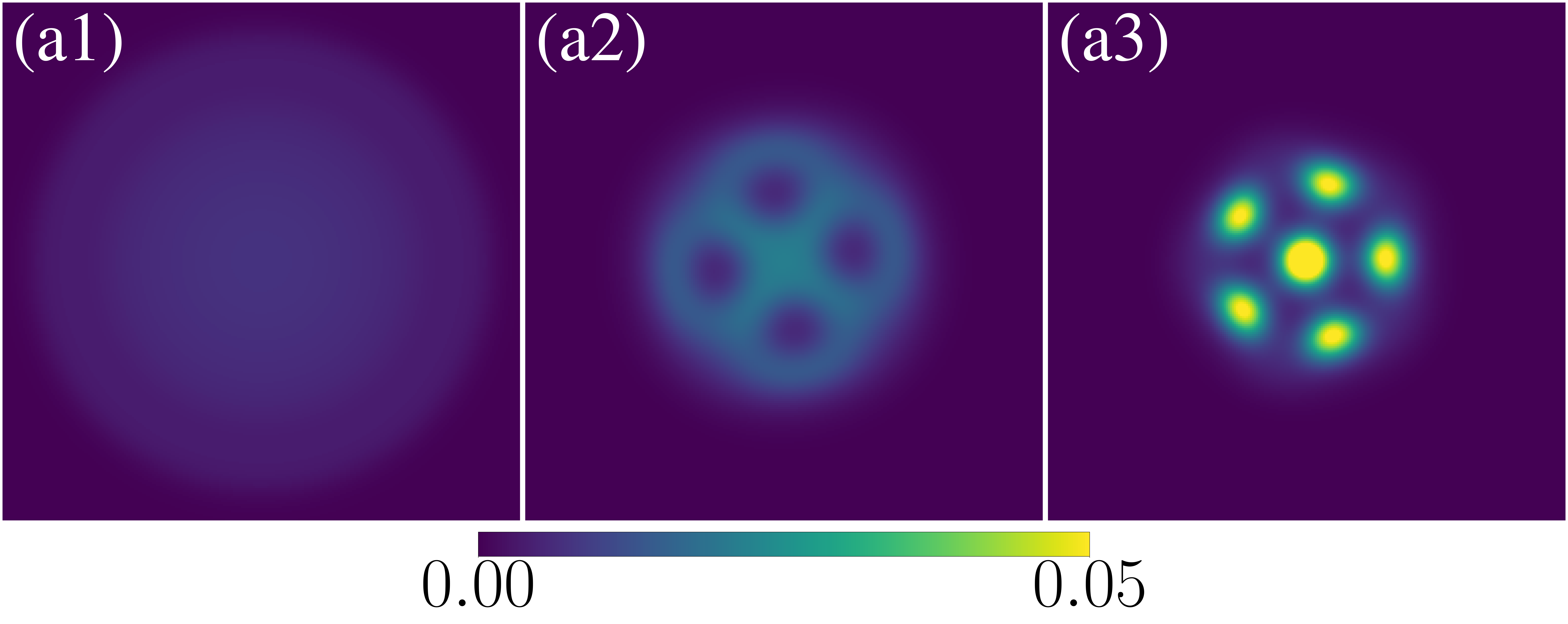}
		\caption{Representative density profiles used to evaluate Leggett's upper bound at $T=50$ nK for flow along $x$: (a1) BEC-like state, $f_S^{\rm (BEC)}\leq0.942$; (a2) connected density-modulated state, $f_S^{\rm (conn.)}\leq0.346$; and (a3) isolated droplet-array state, $f_S^{\rm (drop.)}\leq0.0146$.}
		\label{fig:leggett}
	\end{figure}
	
	We evaluate Eq.~\eqref{eq:leggett_bound} for three representative density profiles: a BEC-like state, a connected density-modulated state, and an isolated droplet-array state. The corresponding bounds are $f_S^{\rm (BEC)}\leq0.942$, $f_S^{\rm (conn.)}\leq0.346$, and $f_S^{\rm (drop.)}\leq0.0146$, as shown in Fig.~\ref{fig:leggett}. The decrease of the bound reflects the increasing inhomogeneity of the density profile. The BEC-like profile gives a value close to unity, while the connected modulated state has a smaller bound because transport is constrained by low-density channels. For the isolated droplet array, the bound is nearly two orders of magnitude smaller than in the BEC-like case, indicating that the density profile strongly suppresses any possible flow along the chosen direction. Since Eq.~\eqref{eq:leggett_bound} gives only an upper bound and depends on the flow direction, these values should be viewed as estimates of geometric constraints rather than direct measurements of the superfluid fraction.
	
	\section{Conclusion}\label{SEC:Conclusion}
	
	We have studied how finite-temperature corrections affect stationary density patterns in a harmonically trapped dipolar Bose--Einstein condensate. Within a temperature-dependent extended Gross--Pitaevskii equation combined with the local-density approximation, the thermal contribution shifts some structural boundaries toward larger scattering lengths and modifies the region in which connected density-modulated states appear. The density contrast shows a sharp change at $T=0$, whereas at finite temperature it varies more gradually, consistent with a smooth crossover-like evolution in the finite trapped system. We also evaluated Leggett's upper bound for representative density profiles, which gives a geometric estimate of the possible superfluid response along a chosen direction.
	
	These results indicate that finite-temperature corrections can play a constructive role in pattern formation in dipolar quantum gases. The structural changes predicted here could be examined through in-situ density imaging, the temperature dependence of the density contrast, and finite-momentum features in static structure factors or time-of-flight images. The relevant control parameters are temperature, scattering length, atom number, and trap geometry. Within the present framework, changing these parameters moves the stationary solution between a nearly uniform condensate and several density-modulated morphologies, including connected droplet-like, honeycomb, and labyrinthine profiles.
	
	The present description is expected to be most reliable for a dilute, weakly depleted gas and within the dense part of the cloud where the local-density treatment of the fluctuation corrections is controlled. Near structural boundaries, in dilute tails, or at larger depletion, coupling to the thermal cloud and higher-order correlations may become important. It would therefore be useful to test the predicted temperature-dependent boundaries using alternative finite-temperature approaches, beyond-LDA treatments, or quantum Monte Carlo calculations. Further studies of cutoff dependence, threshold sensitivity, metastability, and hysteresis under parameter ramps would also help clarify the detailed nature of the smooth evolution seen in the density contrast. Overall, the results suggest that temperature can reshape density-modulated stationary states in dipolar quantum gases and should be regarded as an additional control parameter rather than only as a source of decoherence.
	
	\begin{acknowledgments}
		We acknowledge helpful discussions with Mingyang Guo and Blair Blakie.
		K.-T.X. was supported by the MOST of China (Grant No. G2022181023L) and NUAA (Grant No. YAT22005, No. 2023YJXGG-C32 and No. XCA2405004).
	\end{acknowledgments}

	\appendix
	\section{Derivation of the fitted form}
	\label{appderive}
	The Hamiltonian for a trapped dipolar Bose gas is given by Eq.~(1) of Ref.~\cite{Aybar2019}.
	After splitting the field operator into a condensate part $\Phi(\bm{x})$ and fluctuations $\hat{\varphi}(\bm{x})$,
	and requiring that linear terms in $\hat{\varphi}$ vanish in the many-body ground state, one obtains the generalized GP equation:
	\begin{equation}
		\bigg[ -\frac{\hbar^2 \nabla^2}{2M} - \mu + U_{\text{tr}}(\bm{x}) + \Sigma_H(\bm{x}) + \Delta\mu(\bm{x}) \bigg] \Phi(\bm{x}) = 0,
		\label{eq:13}
	\end{equation}
	where
	\begin{equation}
		\Sigma_H(\bm{x}) = \int d^3x' \, V_{\text{int}}(\bm{x} - \bm{x}') (|\Phi(\bm{x}')|^2 + \tilde{n}(\bm{x}'))
	\end{equation}
	is the Hartree potential (including the direct interaction with both condensate and non-condensate densities), and
	\begin{equation}
		\Delta\mu(\bm{x}) = \Delta_n(\bm{x}) + \Delta_m(\bm{x})
	\end{equation}
	is the fluctuation-induced correction to the chemical potential.
	The terms $\Delta_n$ and $\Delta_m$ involve the non-condensate normal and anomalous densities $\tilde{n}(\bm{x}',\bm{x})$ and $\tilde{m}(\bm{x}',\bm{x})$:
	\begin{align}
		\Delta_n(\bm{x}) \Phi(\bm{x}) &= \int d^3x' \, V_{\text{int}}(\bm{x} - \bm{x}') \, \tilde{n}(\bm{x}',\bm{x}) \, \Phi(\bm{x}'), \\
		\Delta_m(\bm{x}) \Phi(\bm{x}) &= \int d^3x' \, V_{\text{int}}(\bm{x} - \bm{x}') \, \tilde{m}(\bm{x}',\bm{x}) \, \Phi^*(\bm{x}').
	\end{align}

	We assume the condensate density $n_0(\bm{x}) = |\Phi(\bm{x})|^2$ varies slowly on the scale of the wavelength of the elementary excitations.
	The local Bogoliubov problem can then be treated using a WKB-type approximation:
	\begin{align}
		u_j(\bm{x}) &\rightarrow u(\bm{x},\bm{k}) \, e^{i\bm{k}\cdot\bm{x}}, \\
		v_j(\bm{x}) &\rightarrow v(\bm{x},\bm{k}) \, e^{i\bm{k}\cdot\bm{x}}, \\
		E_j &\rightarrow E(\bm{x},\bm{k}), \\
		\sum_j &\rightarrow \int \frac{d^3k}{(2\pi)^3}.
	\end{align}
	Under the LDA, the interaction term simplifies (using the Fourier transform $\tilde{V}_{\text{int}}(\bm{k})$ of $V_{\text{int}}$):
	\begin{align}
		& \Phi(\bm{x}) \int d^3x' V_{\text{int}}(\bm{x} - \bm{x}') \Phi(\bm{x}') u(\bm{x}',\bm{k}) e^{-i\bm{k}\cdot(\bm{x} - \bm{x}')} \nonumber\\
		& \approx n_0(\bm{x}) \, \tilde{V}_{\text{int}}(\bm{k}) \, u(\bm{x},\bm{k}),
	\end{align}
	because $\Phi(\bm{x}') \approx \Phi(\bm{x})$ over the range of $V_{\text{int}}$.
	The same holds for the anomalous term.
	
	The resulting local equations become algebraic:
	\begin{align}
		\varepsilon_k u + n_0 \tilde{V}_{\text{int}} u - n_0 \tilde{V}_{\text{int}} v &= E \, u, \\
		\varepsilon_k v + n_0 \tilde{V}_{\text{int}} v - n_0 \tilde{V}_{\text{int}} u &= -E \, v,
	\end{align}
	with $\varepsilon_k = \frac{\hbar^2 k^2}{2M}$.
	Solving the local equations gives the Bogoliubov dispersion:
	\begin{equation}
		E(\bm{x},k) = \sqrt{\varepsilon_k \bigl( \varepsilon_k + 2 n_0(\bm{x}) \tilde{V}_{\text{int}}(k) \bigr)},
		\label{eq:20}
	\end{equation}
	and the amplitudes:
	\begin{align}
		|v(\bm{x},k)|^2 &= \frac{\varepsilon_k + n_0 \tilde{V}_{\text{int}} - E}{2E}, \\
		u(\bm{x},k) v^*(\bm{x},k) &= \frac{n_0 \tilde{V}_{\text{int}}}{2E}.
		\label{eq:21}
	\end{align}
	These expressions are valid for both zero and finite temperature.

	At $T=0$, the non-condensate densities come only from quantum depletion:
	\begin{align}
		\tilde{n}(\bm{x}) &= \int \frac{d^3k}{(2\pi)^3} |v(\bm{x},k)|^2, \\
		\tilde{m}(\bm{x}) &= -\int \frac{d^3k}{(2\pi)^3} u(\bm{x},k) v^*(\bm{x},k).
	\end{align}
	Substituting \eqref{eq:21} into the definitions of $\Delta_n$ and $\Delta_m$ yields:
	\begin{align}
		\Delta_n(\bm{x}) &= \int \frac{d^3k}{(2\pi)^3} \tilde{V}_{\text{int}}(k) \, |v|^2, \\
		\Delta_m(\bm{x}) &= -\int \frac{d^3k}{(2\pi)^3} \tilde{V}_{\text{int}}(k) \, u v^*.
	\end{align}
	Both integrals are ultraviolet divergent. The standard renormalization subtracts the divergent part corresponding to a free particle gas, leaving a finite result.
	After introducing dimensionless variables:
	\begin{align}
		k &= \frac{q}{\xi(\bm{x})}, \quad \xi(\bm{x}) = \frac{\hbar}{\sqrt{2M g n_0(\bm{x})}}, \\
		u &= \cos\theta_k, \quad f(u) = 1 + \epsilon_{dd} (3u^2 - 1),
	\end{align}
	and imposing a momentum cutoff $k_c = \pi/(2\xi)$ (to exclude the long-wavelength unstable modes for $\epsilon_{dd}>1$), the integrals reduce to
	\begin{equation}
		\Delta_n(\bm{x}) + \Delta_m(\bm{x}) = \frac{32}{3} g \sqrt{\frac{a_s^3}{\pi}} \, Q_5(\epsilon_{dd}) \, |\Phi(\bm{x})|^3,
		\label{eq:23}
	\end{equation}
	where
	\begin{align}
		Q_5(\epsilon_{dd}) = & \frac{1}{4\sqrt{2}} \int_0^1 du \, f(u) \nonumber\\
		& \times \bigg[ \sqrt{\bigg( 4f(u) - q_c^2 \bigg) \bigg( 2f(u) + q_c^2 \bigg)} \nonumber\\
		& \quad - 3f(u)q_c + q_c^3 \bigg], \quad q_c = \frac{\pi}{2}.
	\end{align}
	This gives the local LHY correction for a dipolar gas at $T=0$ in the same cutoff scheme.

	At finite $T$, quasiparticles are thermally excited with Bose occupation numbers:
	\begin{equation}
		\langle \hat{\alpha}_j^\dagger \hat{\alpha}_k \rangle = \delta_{jk} \, N_B(E_j), \quad N_B(E) = \frac{1}{e^{E/k_B T} - 1}.
	\end{equation}
	The normal and anomalous densities acquire extra thermal parts:
	\begin{align}
		\tilde{n}(\bm{x}) &= \int \frac{d^3k}{(2\pi)^3} \Bigl[ |v|^2 + N_B(E) \bigl( |u|^2 + |v|^2 \bigr) \Bigr], \\
		\tilde{m}(\bm{x}) &= -\int \frac{d^3k}{(2\pi)^3} \Bigl[ u v^* + 2 N_B(E) \, u v^* \Bigr].
	\end{align}
	Consequently, the fluctuation correction $\Delta\mu = \Delta_n + \Delta_m$ becomes (after renormalization):
	\begin{align}
		\Delta\mu(\bm{x}) = \int \frac{d^3k}{(2\pi)^3} \tilde{V}_{\text{int}}(k) \bigg[ & \frac{\varepsilon_k}{2E} + \frac{n_0 \tilde{V}_{\text{int}}}{2\varepsilon_k} - \frac12 \nonumber\\
		& + \frac{1}{e^{E/k_B T} - 1} \, \frac{\varepsilon_k}{E} \bigg],
		\label{eq:29}
	\end{align}
	where the first three terms are the quantum (zero-point) contribution (already regularized), and the last term is the thermal contribution.
	
	Using the same dimensionless variables as before and setting $t(\bm{x}) = \frac{k_B T}{g n_0(\bm{x})}$, we obtain:
	\begin{equation}
		\Delta\mu(\bm{x}) = \frac{32}{3} g \sqrt{\frac{a_s^3}{\pi}} \, \bigg[ Q_5(\epsilon_{dd}) + R(\epsilon_{dd}, t(\bm{x})) \bigg] \, |\Phi(\bm{x})|^3,
		\label{eq:32}
	\end{equation}
	with
	\begin{align}
		R(\epsilon_{dd}, t; q_c) = & \frac{3}{4\sqrt{2}} \int_0^1 du 
		\int_{q_c^2}^\infty dQ \, \frac{Q f(u)}{\sqrt{Q+2f(u)}} \nonumber\\
		& \times \frac{1}{e^{\sqrt{Q(Q+2f(u))}/t} - 1}.
		\label{eq:34}
	\end{align}
	Here $Q = q^2$ and the cutoff $q_c = \pi/2$ (spherical cutoff).

	For the parameters considered in the main text, the dimensionless temperature $t=k_BT/(gn_0)$ remains below $10$ in the high-density regions that determine the observed density patterns. \reftable{tab:tem} lists representative values at $T=100\,\text{nK}$. The LDA is not expected to be quantitatively accurate in dilute tails, where $n_0$ becomes very small. The same consideration applies to the $T=200\,\text{nK}$ contrast cuts in Fig.~\ref{C_vs_as}: these curves are used to show how the stationary contrast evolves with temperature in the present model, not to assign high-precision thermodynamic boundaries.
	
	Numerical evaluation of $R(\epsilon_{dd}, t)$ for $0 < t < 10$ shows that it can be very well approximated by a quadratic function in $t$:
	\begin{equation}
		R(\epsilon_{dd}, t) \approx S(\epsilon_{dd}) \, t^2.
	\end{equation}
	The coefficient $S(\epsilon_{dd})$ is obtained by fitting the numerical data. Using the spherical cutoff $k_c = \pi/(2\xi)$, the fit (valid for $0 < \epsilon_{dd} < 2$) is given by:
	\begin{align}
		S(\epsilon_{dd}) = & -0.01029\,\epsilon_{dd}^4 + 0.02963\,\epsilon_{dd}^3 \nonumber\\
		& - 0.05422\,\epsilon_{dd}^2 + 0.009302\,\epsilon_{dd} + 0.1698.
		\label{eq:fit}
	\end{align}
	Now substitute $t^2 = \frac{k_B^2 T^2}{g^2 n_0^2} = \frac{k_B^2 T^2}{g^2 |\Phi(\bm{x})|^4}$ into \eqref{eq:32}:
	\begin{align}
		\Delta\mu(\bm{x}) &= \frac{32}{3} g \sqrt{\frac{a_s^3}{\pi}} Q_5(\epsilon_{dd}) \, |\Phi|^3 + \frac{32}{3} g \sqrt{\frac{a_s^3}{\pi}} S(\epsilon_{dd}) \, \frac{k_B^2 T^2}{g^2 |\Phi|^4} \, |\Phi|^3 \nonumber \\
		&= \frac{32}{3} g \sqrt{\frac{a_s^3}{\pi}} Q_5(\epsilon_{dd}) \, |\Phi|^3 + \frac{32}{3} \sqrt{\frac{a_s^3}{\pi}} \, \frac{S(\epsilon_{dd}) k_B^2}{g} \, \frac{T^2}{|\Phi|}.
	\end{align}
	Thus we define:
	\begin{align}
		\gamma &= \frac{32}{3} g \sqrt{\frac{a_s^3}{\pi}} \, Q_5(\epsilon_{dd}), \\
		\theta &= \frac{32}{3} \sqrt{\frac{a_s^3}{\pi}} \, \frac{S(\epsilon_{dd}) k_B^2}{g}.
	\end{align}
	Finally, the fluctuation chemical potential becomes:
	\begin{equation}
		\Delta\mu(\bm{x}) = \gamma \, |\Phi(\bm{x})|^3 + \theta \, T^2 \, \frac{1}{|\Phi(\bm{x})|}.
		\label{eq:40}
	\end{equation}
	\begin{table}[t]
		\begin{ruledtabular}
			\centering
			\caption{Dimensionless temperature $t=k_BT/(gn_0)$ evaluated at the peak density for representative morphologies at $T=100\,\text{nK}$.}
			\label{tab:tem}
			\begin{tabular}{c c c c c c}
				& BEC  & Modulated & Labyrinth  & Honeycomb & Pumpkin\\
				\midrule
				N($10^3$) & 500 & 10 & 500 & 500 & 800 \\
				$a_s(a_0)$ & 90 & 87 & 87 & 89 & 88.7 \\
				$t$ & 0.4095 & 2.7968 & 0.2727 & 0.3692 & 0.3438 \\
			\end{tabular}
		\end{ruledtabular}	
	\end{table}
    
	We now insert \eqref{eq:40} into \eqref{eq:13}. Within the weak-depletion approximation used in the local-density treatment, the non-condensate density is assumed to be small compared with $|\Phi(\bm{x})|^2$ in the dense region of the cloud. The Hartree potential is then approximated as
	\begin{equation}
		\Sigma_H(\bm{x}) \approx \int d^3x' \, V_{\text{int}}(\bm{x} - \bm{x}') \, |\Phi(\bm{x}')|^2.
	\end{equation}
	Hence the full equation reads:
	\begin{widetext}
		\begin{equation}
			\left[ -\frac{\hbar^2}{2M} \nabla^2 + U_{\text{tr}}(\bm{x}) + \int d^3x' \, V_{\text{int}}(\bm{x} - \bm{x}') |\Phi(\bm{x}')|^2 + \gamma \, |\Phi(\bm{x})|^3 + \theta \, T^2 \frac{1}{|\Phi(\bm{x})|} \right] \Phi(\bm{x}) = \mu \, \Phi(\bm{x}).
			\label{eq:41}
		\end{equation}
	\end{widetext}
	This is the finite-temperature generalized GP equation used here for the high-density region of the condensate.
	The $\gamma$ term originates from zero-point quantum fluctuations (LHY), while the $\theta T^2 / |\Phi|$ term arises from thermal fluctuations and is obtained via the quadratic fit of the function $R(\epsilon_{dd}, t)$.
	
	\bibliography{reference}

\begin{thebibliography}{52}%
\makeatletter
\providecommand \@ifxundefined [1]{%
 \@ifx{#1\undefined}
}%
\providecommand \@ifnum [1]{%
 \ifnum #1\expandafter \@firstoftwo
 \else \expandafter \@secondoftwo
 \fi
}%
\providecommand \@ifx [1]{%
 \ifx #1\expandafter \@firstoftwo
 \else \expandafter \@secondoftwo
 \fi
}%
\providecommand \natexlab [1]{#1}%
\providecommand \enquote  [1]{``#1''}%
\providecommand \bibnamefont  [1]{#1}%
\providecommand \bibfnamefont [1]{#1}%
\providecommand \citenamefont [1]{#1}%
\providecommand \href@noop [0]{\@secondoftwo}%
\providecommand \href [0]{\begingroup \@sanitize@url \@href}%
\providecommand \@href[1]{\@@startlink{#1}\@@href}%
\providecommand \@@href[1]{\endgroup#1\@@endlink}%
\providecommand \@sanitize@url [0]{\catcode `\\12\catcode `\$12\catcode
  `\&12\catcode `\#12\catcode `\^12\catcode `\_12\catcode `\%12\relax}%
\providecommand \@@startlink[1]{}%
\providecommand \@@endlink[0]{}%
\providecommand \url  [0]{\begingroup\@sanitize@url \@url }%
\providecommand \@url [1]{\endgroup\@href {#1}{\urlprefix }}%
\providecommand \urlprefix  [0]{URL }%
\providecommand \Eprint [0]{\href }%
\providecommand \doibase [0]{https://doi.org/}%
\providecommand \selectlanguage [0]{\@gobble}%
\providecommand \bibinfo  [0]{\@secondoftwo}%
\providecommand \bibfield  [0]{\@secondoftwo}%
\providecommand \translation [1]{[#1]}%
\providecommand \BibitemOpen [0]{}%
\providecommand \bibitemStop [0]{}%
\providecommand \bibitemNoStop [0]{.\EOS\space}%
\providecommand \EOS [0]{\spacefactor3000\relax}%
\providecommand \BibitemShut  [1]{\csname bibitem#1\endcsname}%
\let\auto@bib@innerbib\@empty
%</preamble>
\bibitem [{\citenamefont {Lahaye}\ \emph {et~al.}(2009)\citenamefont {Lahaye},
  \citenamefont {Menotti}, \citenamefont {Santos}, \citenamefont {Lewenstein},\
  and\ \citenamefont {Pfau}}]{Lahaye2009}%
  \BibitemOpen
  \bibfield  {author} {\bibinfo {author} {\bibfnamefont {T.}~\bibnamefont
  {Lahaye}}, \bibinfo {author} {\bibfnamefont {C.}~\bibnamefont {Menotti}},
  \bibinfo {author} {\bibfnamefont {L.}~\bibnamefont {Santos}}, \bibinfo
  {author} {\bibfnamefont {M.}~\bibnamefont {Lewenstein}},\ and\ \bibinfo
  {author} {\bibfnamefont {T.}~\bibnamefont {Pfau}},\ }\bibfield  {title}
  {\bibinfo {title} {The physics of dipolar bosonic quantum gases},\ }\href
  {https://doi.org/10.1088/0034-4885/72/12/126401} {\bibfield  {journal}
  {\bibinfo  {journal} {Rep. Prog. Phys.}\ }\textbf {\bibinfo {volume} {72}},\
  \bibinfo {pages} {126401} (\bibinfo {year} {2009})}\BibitemShut {NoStop}%
\bibitem [{\citenamefont {Kora}\ and\ \citenamefont
  {Boninsegni}(2019)}]{YKora2019}%
  \BibitemOpen
  \bibfield  {author} {\bibinfo {author} {\bibfnamefont {Y.}~\bibnamefont
  {Kora}}\ and\ \bibinfo {author} {\bibfnamefont {M.}~\bibnamefont
  {Boninsegni}},\ }\bibfield  {title} {\bibinfo {title} {Patterned supersolids
  in dipolar bose systems},\ }\href
  {https://doi.org/10.1007/s10909-019-02229-z} {\bibfield  {journal} {\bibinfo
  {journal} {J Low Temp Phys.}\ }\textbf {\bibinfo {volume} {197}},\ \bibinfo
  {pages} {337–345} (\bibinfo {year} {2019})}\BibitemShut {NoStop}%
\bibitem [{\citenamefont {B\"ottcher}\ \emph {et~al.}(2021)\citenamefont
  {B\"ottcher}, \citenamefont {Schmidt}, \citenamefont {Hertkorn},
  \citenamefont {Ng}, \citenamefont {Graham}, \citenamefont {Guo},
  \citenamefont {Langen},\ and\ \citenamefont {Pfau}}]{FBottcher2021}%
  \BibitemOpen
  \bibfield  {author} {\bibinfo {author} {\bibfnamefont {F.}~\bibnamefont
  {B\"ottcher}}, \bibinfo {author} {\bibfnamefont {J.-N.}\ \bibnamefont
  {Schmidt}}, \bibinfo {author} {\bibfnamefont {J.}~\bibnamefont {Hertkorn}},
  \bibinfo {author} {\bibfnamefont {K.~S.~H.}\ \bibnamefont {Ng}}, \bibinfo
  {author} {\bibfnamefont {S.~D.}\ \bibnamefont {Graham}}, \bibinfo {author}
  {\bibfnamefont {M.}~\bibnamefont {Guo}}, \bibinfo {author} {\bibfnamefont
  {T.}~\bibnamefont {Langen}},\ and\ \bibinfo {author} {\bibfnamefont
  {T.}~\bibnamefont {Pfau}},\ }\bibfield  {title} {\bibinfo {title} {New states
  of matter with fine-tuned interactions: quantum droplets and dipolar
  supersolids},\ }\href {https://doi.org/10.1088/1361-6633/abc9ab} {\bibfield
  {journal} {\bibinfo  {journal} {Rep. Prog. Phys.}\ }\textbf {\bibinfo
  {volume} {84}},\ \bibinfo {pages} {012403} (\bibinfo {year}
  {2021})}\BibitemShut {NoStop}%
\bibitem [{\citenamefont {Grebenev}\ \emph {et~al.}(1998)\citenamefont
  {Grebenev}, \citenamefont {Toennies},\ and\ \citenamefont
  {Vilesov}}]{Stavropoulos1998}%
  \BibitemOpen
  \bibfield  {author} {\bibinfo {author} {\bibfnamefont {S.}~\bibnamefont
  {Grebenev}}, \bibinfo {author} {\bibfnamefont {J.~P.}\ \bibnamefont
  {Toennies}},\ and\ \bibinfo {author} {\bibfnamefont {A.~F.}\ \bibnamefont
  {Vilesov}},\ }\bibfield  {title} {\bibinfo {title} {Superfluidity within a
  small helium-4 cluster: The microscopic andronikashvili experiment},\ }\href
  {https://doi.org/10.1126/science.279.5359.2083} {\bibfield  {journal}
  {\bibinfo  {journal} {Science}\ }\textbf {\bibinfo {volume} {279}},\ \bibinfo
  {pages} {2083} (\bibinfo {year} {1998})}\BibitemShut {NoStop}%
\bibitem [{\citenamefont {Dalfovo}\ and\ \citenamefont
  {Stringari}(2001)}]{Dalfovo2001}%
  \BibitemOpen
  \bibfield  {author} {\bibinfo {author} {\bibfnamefont {F.}~\bibnamefont
  {Dalfovo}}\ and\ \bibinfo {author} {\bibfnamefont {S.}~\bibnamefont
  {Stringari}},\ }\bibfield  {title} {\bibinfo {title} {Helium nanodroplets and
  trapped bose–einstein condensates as prototypes of finite quantum fluids},\
  }\href {https://doi.org/10.1063/1.1424926} {\bibfield  {journal} {\bibinfo
  {journal} {J. Chem. Phys.}\ }\textbf {\bibinfo {volume} {115}},\ \bibinfo
  {pages} {10078} (\bibinfo {year} {2001})}\BibitemShut {NoStop}%
\bibitem [{\citenamefont {Ravenhall}\ \emph {et~al.}(1983)\citenamefont
  {Ravenhall}, \citenamefont {Pethick},\ and\ \citenamefont
  {Wilson}}]{Ravenhall1983}%
  \BibitemOpen
  \bibfield  {author} {\bibinfo {author} {\bibfnamefont {D.~G.}\ \bibnamefont
  {Ravenhall}}, \bibinfo {author} {\bibfnamefont {C.~J.}\ \bibnamefont
  {Pethick}},\ and\ \bibinfo {author} {\bibfnamefont {J.~R.}\ \bibnamefont
  {Wilson}},\ }\bibfield  {title} {\bibinfo {title} {Structure of matter below
  nuclear saturation density},\ }\href
  {https://doi.org/10.1103/PhysRevLett.50.2066} {\bibfield  {journal} {\bibinfo
   {journal} {Phys. Rev. Lett.}\ }\textbf {\bibinfo {volume} {50}},\ \bibinfo
  {pages} {2066} (\bibinfo {year} {1983})}\BibitemShut {NoStop}%
\bibitem [{\citenamefont {Chamel}\ and\ \citenamefont
  {Haensel}(2008)}]{Chamel2008}%
  \BibitemOpen
  \bibfield  {author} {\bibinfo {author} {\bibfnamefont {N.}~\bibnamefont
  {Chamel}}\ and\ \bibinfo {author} {\bibfnamefont {P.}~\bibnamefont
  {Haensel}},\ }\bibfield  {title} {\bibinfo {title} {Physics of neutron star
  crusts},\ }\href {https://doi.org/10.12942/lrr-2008-10} {\bibfield  {journal}
  {\bibinfo  {journal} {Living Rev. Rel.}\ }\textbf {\bibinfo {volume} {11}},\
  \bibinfo {pages} {10} (\bibinfo {year} {2008})}\BibitemShut {NoStop}%
\bibitem [{\citenamefont {Ferrier-Barbut}\ \emph {et~al.}(2016)\citenamefont
  {Ferrier-Barbut}, \citenamefont {Kadau}, \citenamefont {Schmitt},
  \citenamefont {Wenzel},\ and\ \citenamefont {Pfau}}]{Ferrier-Barbut2016}%
  \BibitemOpen
  \bibfield  {author} {\bibinfo {author} {\bibfnamefont {I.}~\bibnamefont
  {Ferrier-Barbut}}, \bibinfo {author} {\bibfnamefont {H.}~\bibnamefont
  {Kadau}}, \bibinfo {author} {\bibfnamefont {M.}~\bibnamefont {Schmitt}},
  \bibinfo {author} {\bibfnamefont {M.}~\bibnamefont {Wenzel}},\ and\ \bibinfo
  {author} {\bibfnamefont {T.}~\bibnamefont {Pfau}},\ }\bibfield  {title}
  {\bibinfo {title} {Observation of quantum droplets in a strongly dipolar bose
  gas},\ }\href {https://doi.org/10.1103/PhysRevLett.116.215301} {\bibfield
  {journal} {\bibinfo  {journal} {Phys. Rev. Lett.}\ }\textbf {\bibinfo
  {volume} {116}},\ \bibinfo {pages} {215301} (\bibinfo {year}
  {2016})}\BibitemShut {NoStop}%
\bibitem [{\citenamefont {Schmitt}\ \emph {et~al.}(2016)\citenamefont
  {Schmitt}, \citenamefont {Wenzel}, \citenamefont {B\"ottcher}, \citenamefont
  {Ferrier-Barbut},\ and\ \citenamefont {Pfau}}]{Schmitt2016}%
  \BibitemOpen
  \bibfield  {author} {\bibinfo {author} {\bibfnamefont {M.}~\bibnamefont
  {Schmitt}}, \bibinfo {author} {\bibfnamefont {M.}~\bibnamefont {Wenzel}},
  \bibinfo {author} {\bibfnamefont {F.}~\bibnamefont {B\"ottcher}}, \bibinfo
  {author} {\bibfnamefont {I.}~\bibnamefont {Ferrier-Barbut}},\ and\ \bibinfo
  {author} {\bibfnamefont {T.}~\bibnamefont {Pfau}},\ }\bibfield  {title}
  {\bibinfo {title} {Self-bound droplets of a dilute magnetic quantum liquid},\
  }\href {https://doi.org/10.1038/nature20126} {\bibfield  {journal} {\bibinfo
  {journal} {Nature (London)}\ }\textbf {\bibinfo {volume} {539}},\ \bibinfo
  {pages} {259} (\bibinfo {year} {2016})}\BibitemShut {NoStop}%
\bibitem [{\citenamefont {Kadau}\ \emph {et~al.}(2016)\citenamefont {Kadau},
  \citenamefont {Schmitt}, \citenamefont {Wenzel}, \citenamefont {Wink},
  \citenamefont {Maier}, \citenamefont {Ferrier-Barbut},\ and\ \citenamefont
  {Pfau}}]{Kadau2016}%
  \BibitemOpen
  \bibfield  {author} {\bibinfo {author} {\bibfnamefont {H.}~\bibnamefont
  {Kadau}}, \bibinfo {author} {\bibfnamefont {M.}~\bibnamefont {Schmitt}},
  \bibinfo {author} {\bibfnamefont {M.}~\bibnamefont {Wenzel}}, \bibinfo
  {author} {\bibfnamefont {C.}~\bibnamefont {Wink}}, \bibinfo {author}
  {\bibfnamefont {T.}~\bibnamefont {Maier}}, \bibinfo {author} {\bibfnamefont
  {I.}~\bibnamefont {Ferrier-Barbut}},\ and\ \bibinfo {author} {\bibfnamefont
  {T.}~\bibnamefont {Pfau}},\ }\bibfield  {title} {\bibinfo {title} {Quantum
  liquid droplets in a mixture of bose–einstein condensates},\ }\href
  {https://doi.org/10.1038/nature16485} {\bibfield  {journal} {\bibinfo
  {journal} {Nature (London)}\ }\textbf {\bibinfo {volume} {530}},\ \bibinfo
  {pages} {194} (\bibinfo {year} {2016})}\BibitemShut {NoStop}%
\bibitem [{\citenamefont {Tanzi}\ \emph
  {et~al.}(2019{\natexlab{a}})\citenamefont {Tanzi}, \citenamefont {Roccuzzo},
  \citenamefont {Lucioni}, \citenamefont {Fam\`a}, \citenamefont {Fioretti},
  \citenamefont {Gabbanini}, \citenamefont {Modugno}, \citenamefont {Recati},\
  and\ \citenamefont {Stringari}}]{Tanzi2019}%
  \BibitemOpen
  \bibfield  {author} {\bibinfo {author} {\bibfnamefont {L.}~\bibnamefont
  {Tanzi}}, \bibinfo {author} {\bibfnamefont {S.~M.}\ \bibnamefont {Roccuzzo}},
  \bibinfo {author} {\bibfnamefont {E.}~\bibnamefont {Lucioni}}, \bibinfo
  {author} {\bibfnamefont {F.}~\bibnamefont {Fam\`a}}, \bibinfo {author}
  {\bibfnamefont {A.}~\bibnamefont {Fioretti}}, \bibinfo {author}
  {\bibfnamefont {C.}~\bibnamefont {Gabbanini}}, \bibinfo {author}
  {\bibfnamefont {G.}~\bibnamefont {Modugno}}, \bibinfo {author} {\bibfnamefont
  {A.}~\bibnamefont {Recati}},\ and\ \bibinfo {author} {\bibfnamefont
  {S.}~\bibnamefont {Stringari}},\ }\bibfield  {title} {\bibinfo {title}
  {Supersolid symmetry breaking from compressional oscillations in a dipolar
  quantum gas},\ }\href {https://doi.org/10.1038/s41586-019-1568-6} {\bibfield
  {journal} {\bibinfo  {journal} {Nature (London)}\ }\textbf {\bibinfo {volume}
  {574}},\ \bibinfo {pages} {382} (\bibinfo {year}
  {2019}{\natexlab{a}})}\BibitemShut {NoStop}%
\bibitem [{\citenamefont {Guo}\ \emph {et~al.}(2019)\citenamefont {Guo},
  \citenamefont {B\"ottcher}, \citenamefont {Hertkorn}, \citenamefont
  {Schmidt}, \citenamefont {Wenzel}, \citenamefont {B\"uchler}, \citenamefont
  {Langen},\ and\ \citenamefont {Pfau}}]{Guo2019}%
  \BibitemOpen
  \bibfield  {author} {\bibinfo {author} {\bibfnamefont {M.}~\bibnamefont
  {Guo}}, \bibinfo {author} {\bibfnamefont {F.}~\bibnamefont {B\"ottcher}},
  \bibinfo {author} {\bibfnamefont {J.}~\bibnamefont {Hertkorn}}, \bibinfo
  {author} {\bibfnamefont {J.-N.}\ \bibnamefont {Schmidt}}, \bibinfo {author}
  {\bibfnamefont {M.}~\bibnamefont {Wenzel}}, \bibinfo {author} {\bibfnamefont
  {H.~P.}\ \bibnamefont {B\"uchler}}, \bibinfo {author} {\bibfnamefont
  {T.}~\bibnamefont {Langen}},\ and\ \bibinfo {author} {\bibfnamefont
  {T.}~\bibnamefont {Pfau}},\ }\bibfield  {title} {\bibinfo {title} {The
  low-energy goldstone mode in a trapped dipolar supersolid},\ }\href
  {https://doi.org/10.1038/s41586-019-1569-5} {\bibfield  {journal} {\bibinfo
  {journal} {Nature (London)}\ }\textbf {\bibinfo {volume} {574}},\ \bibinfo
  {pages} {386} (\bibinfo {year} {2019})}\BibitemShut {NoStop}%
\bibitem [{\citenamefont {Natale}\ \emph {et~al.}(2019)\citenamefont {Natale},
  \citenamefont {van Bijnen}, \citenamefont {Patscheider}, \citenamefont
  {Petter}, \citenamefont {Mark}, \citenamefont {Chomaz},\ and\ \citenamefont
  {Ferlaino}}]{Natale2019}%
  \BibitemOpen
  \bibfield  {author} {\bibinfo {author} {\bibfnamefont {G.}~\bibnamefont
  {Natale}}, \bibinfo {author} {\bibfnamefont {R.~M.~W.}\ \bibnamefont {van
  Bijnen}}, \bibinfo {author} {\bibfnamefont {A.}~\bibnamefont {Patscheider}},
  \bibinfo {author} {\bibfnamefont {D.}~\bibnamefont {Petter}}, \bibinfo
  {author} {\bibfnamefont {M.~J.}\ \bibnamefont {Mark}}, \bibinfo {author}
  {\bibfnamefont {L.}~\bibnamefont {Chomaz}},\ and\ \bibinfo {author}
  {\bibfnamefont {F.}~\bibnamefont {Ferlaino}},\ }\bibfield  {title} {\bibinfo
  {title} {Excitation spectrum of a trapped dipolar supersolid and its
  experimental evidence},\ }\href
  {https://doi.org/10.1103/PhysRevLett.123.050402} {\bibfield  {journal}
  {\bibinfo  {journal} {Phys. Rev. Lett.}\ }\textbf {\bibinfo {volume} {123}},\
  \bibinfo {pages} {050402} (\bibinfo {year} {2019})}\BibitemShut {NoStop}%
\bibitem [{\citenamefont {B\"ottcher}\ \emph {et~al.}(2019)\citenamefont
  {B\"ottcher}, \citenamefont {Schmidt}, \citenamefont {Wenzel}, \citenamefont
  {Hertkorn}, \citenamefont {Guo}, \citenamefont {Langen},\ and\ \citenamefont
  {Pfau}}]{Bottcher2019}%
  \BibitemOpen
  \bibfield  {author} {\bibinfo {author} {\bibfnamefont {F.}~\bibnamefont
  {B\"ottcher}}, \bibinfo {author} {\bibfnamefont {J.-N.}\ \bibnamefont
  {Schmidt}}, \bibinfo {author} {\bibfnamefont {M.}~\bibnamefont {Wenzel}},
  \bibinfo {author} {\bibfnamefont {J.}~\bibnamefont {Hertkorn}}, \bibinfo
  {author} {\bibfnamefont {M.}~\bibnamefont {Guo}}, \bibinfo {author}
  {\bibfnamefont {T.}~\bibnamefont {Langen}},\ and\ \bibinfo {author}
  {\bibfnamefont {T.}~\bibnamefont {Pfau}},\ }\bibfield  {title} {\bibinfo
  {title} {Transient supersolid properties in an array of dipolar quantum
  droplets},\ }\href {https://doi.org/10.1103/PhysRevX.9.011051} {\bibfield
  {journal} {\bibinfo  {journal} {Phys. Rev. X}\ }\textbf {\bibinfo {volume}
  {9}},\ \bibinfo {pages} {011051} (\bibinfo {year} {2019})}\BibitemShut
  {NoStop}%
\bibitem [{\citenamefont {Chomaz}\ \emph {et~al.}(2018)\citenamefont {Chomaz},
  \citenamefont {van Bijnen}, \citenamefont {Petter}, \citenamefont {Faraoni},
  \citenamefont {Baier}, \citenamefont {Becher}, \citenamefont {Mark},
  \citenamefont {Wächtler}, \citenamefont {Santos},\ and\ \citenamefont
  {Ferlaino}}]{Chomaz2018}%
  \BibitemOpen
  \bibfield  {author} {\bibinfo {author} {\bibfnamefont {L.}~\bibnamefont
  {Chomaz}}, \bibinfo {author} {\bibfnamefont {R.~M.~W.}\ \bibnamefont {van
  Bijnen}}, \bibinfo {author} {\bibfnamefont {D.}~\bibnamefont {Petter}},
  \bibinfo {author} {\bibfnamefont {G.}~\bibnamefont {Faraoni}}, \bibinfo
  {author} {\bibfnamefont {S.}~\bibnamefont {Baier}}, \bibinfo {author}
  {\bibfnamefont {J.~H.}\ \bibnamefont {Becher}}, \bibinfo {author}
  {\bibfnamefont {M.~J.}\ \bibnamefont {Mark}}, \bibinfo {author}
  {\bibfnamefont {F.}~\bibnamefont {Wächtler}}, \bibinfo {author}
  {\bibfnamefont {L.}~\bibnamefont {Santos}},\ and\ \bibinfo {author}
  {\bibfnamefont {F.}~\bibnamefont {Ferlaino}},\ }\bibfield  {title} {\bibinfo
  {title} {Observation of roton mode population in a dipolar quantum gas},\
  }\href {https://doi.org/10.1038/s41567-018-0054-7} {\bibfield  {journal}
  {\bibinfo  {journal} {Nat. Phys.}\ }\textbf {\bibinfo {volume} {14}},\
  \bibinfo {pages} {442} (\bibinfo {year} {2018})}\BibitemShut {NoStop}%
\bibitem [{\citenamefont {Chomaz}\ \emph {et~al.}(2019)\citenamefont {Chomaz},
  \citenamefont {Petter}, \citenamefont {Ilzh\"ofer}, \citenamefont {Natale},
  \citenamefont {Trautmann}, \citenamefont {Politi}, \citenamefont
  {Durastante}, \citenamefont {van Bijnen}, \citenamefont {Patscheider},
  \citenamefont {Sohmen}, \citenamefont {Mark},\ and\ \citenamefont
  {Ferlaino}}]{Chomaz2019}%
  \BibitemOpen
  \bibfield  {author} {\bibinfo {author} {\bibfnamefont {L.}~\bibnamefont
  {Chomaz}}, \bibinfo {author} {\bibfnamefont {D.}~\bibnamefont {Petter}},
  \bibinfo {author} {\bibfnamefont {P.}~\bibnamefont {Ilzh\"ofer}}, \bibinfo
  {author} {\bibfnamefont {G.}~\bibnamefont {Natale}}, \bibinfo {author}
  {\bibfnamefont {A.}~\bibnamefont {Trautmann}}, \bibinfo {author}
  {\bibfnamefont {C.}~\bibnamefont {Politi}}, \bibinfo {author} {\bibfnamefont
  {G.}~\bibnamefont {Durastante}}, \bibinfo {author} {\bibfnamefont {R.~M.~W.}\
  \bibnamefont {van Bijnen}}, \bibinfo {author} {\bibfnamefont
  {A.}~\bibnamefont {Patscheider}}, \bibinfo {author} {\bibfnamefont
  {M.}~\bibnamefont {Sohmen}}, \bibinfo {author} {\bibfnamefont {M.~J.}\
  \bibnamefont {Mark}},\ and\ \bibinfo {author} {\bibfnamefont
  {F.}~\bibnamefont {Ferlaino}},\ }\bibfield  {title} {\bibinfo {title}
  {Long-lived and transient supersolid behaviors in dipolar quantum gases},\
  }\href {https://doi.org/10.1103/PhysRevX.9.021012} {\bibfield  {journal}
  {\bibinfo  {journal} {Phys. Rev. X}\ }\textbf {\bibinfo {volume} {9}},\
  \bibinfo {pages} {021012} (\bibinfo {year} {2019})}\BibitemShut {NoStop}%
\bibitem [{\citenamefont {Ilzh\"ofer}\ \emph {et~al.}(2021)\citenamefont
  {Ilzh\"ofer}, \citenamefont {Sohmen}, \citenamefont {Durastante},
  \citenamefont {Politi}, \citenamefont {Trautmann}, \citenamefont {Natale},
  \citenamefont {Morpurgo}, \citenamefont {Giamarchi}, \citenamefont {Chomaz},
  \citenamefont {Mark},\ and\ \citenamefont {Ferlaino}}]{Ilzhofer2021}%
  \BibitemOpen
  \bibfield  {author} {\bibinfo {author} {\bibfnamefont {P.}~\bibnamefont
  {Ilzh\"ofer}}, \bibinfo {author} {\bibfnamefont {M.}~\bibnamefont {Sohmen}},
  \bibinfo {author} {\bibfnamefont {G.}~\bibnamefont {Durastante}}, \bibinfo
  {author} {\bibfnamefont {C.}~\bibnamefont {Politi}}, \bibinfo {author}
  {\bibfnamefont {A.}~\bibnamefont {Trautmann}}, \bibinfo {author}
  {\bibfnamefont {G.}~\bibnamefont {Natale}}, \bibinfo {author} {\bibfnamefont
  {G.}~\bibnamefont {Morpurgo}}, \bibinfo {author} {\bibfnamefont
  {T.}~\bibnamefont {Giamarchi}}, \bibinfo {author} {\bibfnamefont
  {L.}~\bibnamefont {Chomaz}}, \bibinfo {author} {\bibfnamefont {M.~J.}\
  \bibnamefont {Mark}},\ and\ \bibinfo {author} {\bibfnamefont
  {F.}~\bibnamefont {Ferlaino}},\ }\bibfield  {title} {\bibinfo {title} {Phase
  coherence in out-of-equilibrium supersolid states of ultracold dipolar
  atoms},\ }\href {https://doi.org/10.1038/s41567-020-01100-3} {\bibfield
  {journal} {\bibinfo  {journal} {Nat. Phys.}\ }\textbf {\bibinfo {volume}
  {17}},\ \bibinfo {pages} {356} (\bibinfo {year} {2021})}\BibitemShut
  {NoStop}%
\bibitem [{\citenamefont {Hertkorn}\ \emph {et~al.}(2021)\citenamefont
  {Hertkorn}, \citenamefont {Schmidt}, \citenamefont {Guo}, \citenamefont
  {B\"ottcher}, \citenamefont {Ng}, \citenamefont {Graham}, \citenamefont
  {Uerlings}, \citenamefont {Langen}, \citenamefont {Zwierlein},\ and\
  \citenamefont {Pfau}}]{Hertkorn2021}%
  \BibitemOpen
  \bibfield  {author} {\bibinfo {author} {\bibfnamefont {J.}~\bibnamefont
  {Hertkorn}}, \bibinfo {author} {\bibfnamefont {J.-N.}\ \bibnamefont
  {Schmidt}}, \bibinfo {author} {\bibfnamefont {M.}~\bibnamefont {Guo}},
  \bibinfo {author} {\bibfnamefont {F.}~\bibnamefont {B\"ottcher}}, \bibinfo
  {author} {\bibfnamefont {K.~S.~H.}\ \bibnamefont {Ng}}, \bibinfo {author}
  {\bibfnamefont {S.~D.}\ \bibnamefont {Graham}}, \bibinfo {author}
  {\bibfnamefont {P.}~\bibnamefont {Uerlings}}, \bibinfo {author}
  {\bibfnamefont {T.}~\bibnamefont {Langen}}, \bibinfo {author} {\bibfnamefont
  {M.}~\bibnamefont {Zwierlein}},\ and\ \bibinfo {author} {\bibfnamefont
  {T.}~\bibnamefont {Pfau}},\ }\bibfield  {title} {\bibinfo {title} {Pattern
  formation in quantum ferrofluids: From supersolids to superglasses},\ }\href
  {https://doi.org/10.1103/PhysRevResearch.3.033125} {\bibfield  {journal}
  {\bibinfo  {journal} {Phys. Rev. Res.}\ }\textbf {\bibinfo {volume} {3}},\
  \bibinfo {pages} {033125} (\bibinfo {year} {2021})}\BibitemShut {NoStop}%
\bibitem [{\citenamefont {Zhang}\ \emph {et~al.}(2021)\citenamefont {Zhang},
  \citenamefont {Pohl},\ and\ \citenamefont {Maucher}}]{Zhang2021}%
  \BibitemOpen
  \bibfield  {author} {\bibinfo {author} {\bibfnamefont {Y.-C.}\ \bibnamefont
  {Zhang}}, \bibinfo {author} {\bibfnamefont {T.}~\bibnamefont {Pohl}},\ and\
  \bibinfo {author} {\bibfnamefont {F.}~\bibnamefont {Maucher}},\ }\bibfield
  {title} {\bibinfo {title} {Phases of supersolids in confined dipolar
  bose-einstein condensates},\ }\href
  {https://doi.org/10.1103/PhysRevA.104.013310} {\bibfield  {journal} {\bibinfo
   {journal} {Phys. Rev. A}\ }\textbf {\bibinfo {volume} {104}},\ \bibinfo
  {pages} {013310} (\bibinfo {year} {2021})}\BibitemShut {NoStop}%
\bibitem [{\citenamefont {Tanzi}\ \emph
  {et~al.}(2019{\natexlab{b}})\citenamefont {Tanzi}, \citenamefont {Lucioni},
  \citenamefont {Fam\`a}, \citenamefont {Catani}, \citenamefont {Fioretti},
  \citenamefont {Gabbanini}, \citenamefont {Bisset}, \citenamefont {Santos},\
  and\ \citenamefont {Modugno}}]{Tanzi2019prl}%
  \BibitemOpen
  \bibfield  {author} {\bibinfo {author} {\bibfnamefont {L.}~\bibnamefont
  {Tanzi}}, \bibinfo {author} {\bibfnamefont {E.}~\bibnamefont {Lucioni}},
  \bibinfo {author} {\bibfnamefont {F.}~\bibnamefont {Fam\`a}}, \bibinfo
  {author} {\bibfnamefont {J.}~\bibnamefont {Catani}}, \bibinfo {author}
  {\bibfnamefont {A.}~\bibnamefont {Fioretti}}, \bibinfo {author}
  {\bibfnamefont {C.}~\bibnamefont {Gabbanini}}, \bibinfo {author}
  {\bibfnamefont {R.~N.}\ \bibnamefont {Bisset}}, \bibinfo {author}
  {\bibfnamefont {L.}~\bibnamefont {Santos}},\ and\ \bibinfo {author}
  {\bibfnamefont {G.}~\bibnamefont {Modugno}},\ }\bibfield  {title} {\bibinfo
  {title} {Observation of a dipolar quantum gas with metastable supersolid
  properties},\ }\href {https://doi.org/10.1103/PhysRevLett.122.130405}
  {\bibfield  {journal} {\bibinfo  {journal} {Phys. Rev. Lett.}\ }\textbf
  {\bibinfo {volume} {122}},\ \bibinfo {pages} {130405} (\bibinfo {year}
  {2019}{\natexlab{b}})}\BibitemShut {NoStop}%
\bibitem [{\citenamefont {Norcia}\ \emph {et~al.}(2021)\citenamefont {Norcia},
  \citenamefont {Politi}, \citenamefont {Klaus}, \citenamefont {Poli},
  \citenamefont {Sohmen}, \citenamefont {Mark}, \citenamefont {Bisset},
  \citenamefont {Santos},\ and\ \citenamefont {Ferlaino}}]{Norcia2021}%
  \BibitemOpen
  \bibfield  {author} {\bibinfo {author} {\bibfnamefont {M.~A.}\ \bibnamefont
  {Norcia}}, \bibinfo {author} {\bibfnamefont {C.}~\bibnamefont {Politi}},
  \bibinfo {author} {\bibfnamefont {L.}~\bibnamefont {Klaus}}, \bibinfo
  {author} {\bibfnamefont {E.}~\bibnamefont {Poli}}, \bibinfo {author}
  {\bibfnamefont {M.}~\bibnamefont {Sohmen}}, \bibinfo {author} {\bibfnamefont
  {M.~J.}\ \bibnamefont {Mark}}, \bibinfo {author} {\bibfnamefont {R.~N.}\
  \bibnamefont {Bisset}}, \bibinfo {author} {\bibfnamefont {L.}~\bibnamefont
  {Santos}},\ and\ \bibinfo {author} {\bibfnamefont {F.}~\bibnamefont
  {Ferlaino}},\ }\bibfield  {title} {\bibinfo {title} {Two-dimensional
  supersolidity in a dipolar quantum gas},\ }\href
  {https://doi.org/10.1038/s41586-021-03725-7} {\bibfield  {journal} {\bibinfo
  {journal} {Nature (London)}\ }\textbf {\bibinfo {volume} {596}},\ \bibinfo
  {pages} {357} (\bibinfo {year} {2021})}\BibitemShut {NoStop}%
\bibitem [{\citenamefont {Lima}\ and\ \citenamefont
  {Pelster}(2011)}]{Lima2011}%
  \BibitemOpen
  \bibfield  {author} {\bibinfo {author} {\bibfnamefont {A.~R.~P.}\
  \bibnamefont {Lima}}\ and\ \bibinfo {author} {\bibfnamefont {A.}~\bibnamefont
  {Pelster}},\ }\bibfield  {title} {\bibinfo {title} {Quantum fluctuations in
  dipolar bose gases},\ }\href {https://doi.org/10.1103/PhysRevA.84.041604}
  {\bibfield  {journal} {\bibinfo  {journal} {Phys. Rev. A}\ }\textbf {\bibinfo
  {volume} {84}},\ \bibinfo {pages} {041604} (\bibinfo {year}
  {2011})}\BibitemShut {NoStop}%
\bibitem [{\citenamefont {Lima}\ and\ \citenamefont
  {Pelster}(2012)}]{Lima2012}%
  \BibitemOpen
  \bibfield  {author} {\bibinfo {author} {\bibfnamefont {A.~R.~P.}\
  \bibnamefont {Lima}}\ and\ \bibinfo {author} {\bibfnamefont {A.}~\bibnamefont
  {Pelster}},\ }\bibfield  {title} {\bibinfo {title} {Beyond mean-field
  low-lying excitations of dipolar bose gases},\ }\href
  {https://doi.org/10.1103/PhysRevA.86.063609} {\bibfield  {journal} {\bibinfo
  {journal} {Phys. Rev. A}\ }\textbf {\bibinfo {volume} {86}},\ \bibinfo
  {pages} {063609} (\bibinfo {year} {2012})}\BibitemShut {NoStop}%
\bibitem [{\citenamefont {Seul}\ and\ \citenamefont
  {Andelman}(1995)}]{Seul1995}%
  \BibitemOpen
  \bibfield  {author} {\bibinfo {author} {\bibfnamefont {M.}~\bibnamefont
  {Seul}}\ and\ \bibinfo {author} {\bibfnamefont {D.}~\bibnamefont
  {Andelman}},\ }\bibfield  {title} {\bibinfo {title} {Domain shapes and
  patterns: The phenomenology of modulated phases},\ }\href
  {https://doi.org/10.1126/science.267.5197.476} {\bibfield  {journal}
  {\bibinfo  {journal} {Science}\ }\textbf {\bibinfo {volume} {267}},\ \bibinfo
  {pages} {476} (\bibinfo {year} {1995})}\BibitemShut {NoStop}%
\bibitem [{\citenamefont {Rosensweig}(1997)}]{Rosensweig1997}%
  \BibitemOpen
  \bibfield  {author} {\bibinfo {author} {\bibfnamefont {R.~E.}\ \bibnamefont
  {Rosensweig}},\ }\href@noop {} {\emph {\bibinfo {title} {Ferrohydrodynamics,
  Dover Books on Physics}}}\ (\bibinfo  {publisher} {Dover},\ \bibinfo
  {address} {New York},\ \bibinfo {year} {1997})\BibitemShut {NoStop}%
\bibitem [{\citenamefont {Andelman}\ and\ \citenamefont
  {Ravenhall}(2009)}]{Andelman2009}%
  \BibitemOpen
  \bibfield  {author} {\bibinfo {author} {\bibfnamefont {D.}~\bibnamefont
  {Andelman}}\ and\ \bibinfo {author} {\bibfnamefont {R.~E.}\ \bibnamefont
  {Ravenhall}},\ }\bibfield  {title} {\bibinfo {title} {Modulated phases:
  Review and recent results},\ }\href {https://doi.org/10.1021/jp807770n}
  {\bibfield  {journal} {\bibinfo  {journal} {J. Phys. Chem. B}\ }\textbf
  {\bibinfo {volume} {113}},\ \bibinfo {pages} {3785} (\bibinfo {year}
  {2009})}\BibitemShut {NoStop}%
\bibitem [{\citenamefont {Bourgine}\ and\ \citenamefont
  {Lesne}(2011)}]{Bourgine2011}%
  \BibitemOpen
  \bibfield  {author} {\bibinfo {author} {\bibfnamefont {P.}~\bibnamefont
  {Bourgine}}\ and\ \bibinfo {author} {\bibfnamefont {A.}~\bibnamefont
  {Lesne}},\ }\href@noop {} {\emph {\bibinfo {title} {Morphogenesis, Origins of
  Pat- terns and Shapes, Springer Complexity}}}\ (\bibinfo  {publisher}
  {Springer, Berlin, Heidelberg},\ \bibinfo {year} {2011})\BibitemShut
  {NoStop}%
\bibitem [{\citenamefont {Zhang}\ \emph {et~al.}(2024)\citenamefont {Zhang},
  \citenamefont {Pohl},\ and\ \citenamefont {Maucher}}]{Zhang2024}%
  \BibitemOpen
  \bibfield  {author} {\bibinfo {author} {\bibfnamefont {Y.-C.}\ \bibnamefont
  {Zhang}}, \bibinfo {author} {\bibfnamefont {T.}~\bibnamefont {Pohl}},\ and\
  \bibinfo {author} {\bibfnamefont {F.}~\bibnamefont {Maucher}},\ }\bibfield
  {title} {\bibinfo {title} {Metastable patterns in one- and two-component
  dipolar bose-einstein condensates},\ }\href
  {https://doi.org/10.1103/PhysRevResearch.6.023023} {\bibfield  {journal}
  {\bibinfo  {journal} {Phys. Rev. Res.}\ }\textbf {\bibinfo {volume} {6}},\
  \bibinfo {pages} {023023} (\bibinfo {year} {2024})}\BibitemShut {NoStop}%
\bibitem [{\citenamefont {Griffin}(1996)}]{Griffin1996}%
  \BibitemOpen
  \bibfield  {author} {\bibinfo {author} {\bibfnamefont {A.}~\bibnamefont
  {Griffin}},\ }\bibfield  {title} {\bibinfo {title} {Conserving and gapless
  approximations for an inhomogeneous bose gas at finite temperatures},\ }\href
  {https://doi.org/10.1103/PhysRevB.53.9341} {\bibfield  {journal} {\bibinfo
  {journal} {Phys. Rev. B}\ }\textbf {\bibinfo {volume} {53}},\ \bibinfo
  {pages} {9341} (\bibinfo {year} {1996})}\BibitemShut {NoStop}%
\bibitem [{\citenamefont {Ronen}\ and\ \citenamefont {Bohn}(2007)}]{Ronen2007}%
  \BibitemOpen
  \bibfield  {author} {\bibinfo {author} {\bibfnamefont {S.}~\bibnamefont
  {Ronen}}\ and\ \bibinfo {author} {\bibfnamefont {J.~L.}\ \bibnamefont
  {Bohn}},\ }\bibfield  {title} {\bibinfo {title} {Dipolar bose-einstein
  condensates at finite temperature},\ }\href
  {https://doi.org/10.1103/PhysRevA.76.043607} {\bibfield  {journal} {\bibinfo
  {journal} {Phys. Rev. A}\ }\textbf {\bibinfo {volume} {76}},\ \bibinfo
  {pages} {043607} (\bibinfo {year} {2007})}\BibitemShut {NoStop}%
\bibitem [{\citenamefont {Bisset}\ \emph {et~al.}(2012)\citenamefont {Bisset},
  \citenamefont {Baillie},\ and\ \citenamefont {Blakie}}]{Bisset2012}%
  \BibitemOpen
  \bibfield  {author} {\bibinfo {author} {\bibfnamefont {R.~N.}\ \bibnamefont
  {Bisset}}, \bibinfo {author} {\bibfnamefont {D.}~\bibnamefont {Baillie}},\
  and\ \bibinfo {author} {\bibfnamefont {P.~B.}\ \bibnamefont {Blakie}},\
  }\bibfield  {title} {\bibinfo {title} {Finite-temperature trapped dipolar
  bose gas},\ }\href {https://doi.org/10.1103/PhysRevA.86.033609} {\bibfield
  {journal} {\bibinfo  {journal} {Phys. Rev. A}\ }\textbf {\bibinfo {volume}
  {86}},\ \bibinfo {pages} {033609} (\bibinfo {year} {2012})}\BibitemShut
  {NoStop}%
\bibitem [{\citenamefont {Ticknor}(2012)}]{Ticknor2012}%
  \BibitemOpen
  \bibfield  {author} {\bibinfo {author} {\bibfnamefont {C.}~\bibnamefont
  {Ticknor}},\ }\bibfield  {title} {\bibinfo {title} {Finite-temperature
  analysis of a quasi-two-dimensional dipolar gas},\ }\href
  {https://doi.org/10.1103/PhysRevA.85.033629} {\bibfield  {journal} {\bibinfo
  {journal} {Phys. Rev. A}\ }\textbf {\bibinfo {volume} {85}},\ \bibinfo
  {pages} {033629} (\bibinfo {year} {2012})}\BibitemShut {NoStop}%
\bibitem [{\citenamefont {Paw\l{}owski}\ \emph {et~al.}(2013)\citenamefont
  {Paw\l{}owski}, \citenamefont {Bienias}, \citenamefont {Pfau},\ and\
  \citenamefont {Rz\k{a}\ifmmode~\dot{z}\else
  \.{z}\fi{}ewski}}]{Pawlowski2013}%
  \BibitemOpen
  \bibfield  {author} {\bibinfo {author} {\bibfnamefont {K.}~\bibnamefont
  {Paw\l{}owski}}, \bibinfo {author} {\bibfnamefont {P.}~\bibnamefont
  {Bienias}}, \bibinfo {author} {\bibfnamefont {T.}~\bibnamefont {Pfau}},\ and\
  \bibinfo {author} {\bibfnamefont {K.}~\bibnamefont
  {Rz\k{a}\ifmmode~\dot{z}\else \.{z}\fi{}ewski}},\ }\bibfield  {title}
  {\bibinfo {title} {Correlations of a quasi-two-dimensional dipolar ultracold
  gas at finite temperatures},\ }\href
  {https://doi.org/10.1103/PhysRevA.87.043620} {\bibfield  {journal} {\bibinfo
  {journal} {Phys. Rev. A}\ }\textbf {\bibinfo {volume} {87}},\ \bibinfo
  {pages} {043620} (\bibinfo {year} {2013})}\BibitemShut {NoStop}%
\bibitem [{\citenamefont {Aybar}\ and\ \citenamefont
  {Oktel}(2019)}]{Aybar2019}%
  \BibitemOpen
  \bibfield  {author} {\bibinfo {author} {\bibfnamefont {E.}~\bibnamefont
  {Aybar}}\ and\ \bibinfo {author} {\bibfnamefont {M.~O.}\ \bibnamefont
  {Oktel}},\ }\bibfield  {title} {\bibinfo {title} {Temperature-dependent
  density profiles of dipolar droplets},\ }\href
  {https://doi.org/10.1103/PhysRevA.99.013620} {\bibfield  {journal} {\bibinfo
  {journal} {Phys. Rev. A}\ }\textbf {\bibinfo {volume} {99}},\ \bibinfo
  {pages} {013620} (\bibinfo {year} {2019})}\BibitemShut {NoStop}%
\bibitem [{\citenamefont {\"Ozt\"urk}\ \emph {et~al.}(2020)\citenamefont
  {\"Ozt\"urk}, \citenamefont {Aybar},\ and\ \citenamefont
  {Oktel}}]{Ozturk2020}%
  \BibitemOpen
  \bibfield  {author} {\bibinfo {author} {\bibfnamefont {S.~F.}\ \bibnamefont
  {\"Ozt\"urk}}, \bibinfo {author} {\bibfnamefont {E.}~\bibnamefont {Aybar}},\
  and\ \bibinfo {author} {\bibfnamefont {M.~O.}\ \bibnamefont {Oktel}},\
  }\bibfield  {title} {\bibinfo {title} {Temperature dependence of the density
  and excitations of dipolar droplets},\ }\href
  {https://doi.org/10.1103/PhysRevA.102.033329} {\bibfield  {journal} {\bibinfo
   {journal} {Phys. Rev. A}\ }\textbf {\bibinfo {volume} {102}},\ \bibinfo
  {pages} {033329} (\bibinfo {year} {2020})}\BibitemShut {NoStop}%
\bibitem [{\citenamefont {Sohmen}\ \emph {et~al.}(2021)\citenamefont {Sohmen},
  \citenamefont {Politi}, \citenamefont {Klaus}, \citenamefont {Chomaz},
  \citenamefont {Mark}, \citenamefont {Norcia},\ and\ \citenamefont
  {Ferlaino}}]{Sohmen2021}%
  \BibitemOpen
  \bibfield  {author} {\bibinfo {author} {\bibfnamefont {M.}~\bibnamefont
  {Sohmen}}, \bibinfo {author} {\bibfnamefont {C.}~\bibnamefont {Politi}},
  \bibinfo {author} {\bibfnamefont {L.}~\bibnamefont {Klaus}}, \bibinfo
  {author} {\bibfnamefont {L.}~\bibnamefont {Chomaz}}, \bibinfo {author}
  {\bibfnamefont {M.~J.}\ \bibnamefont {Mark}}, \bibinfo {author}
  {\bibfnamefont {M.~A.}\ \bibnamefont {Norcia}},\ and\ \bibinfo {author}
  {\bibfnamefont {F.}~\bibnamefont {Ferlaino}},\ }\bibfield  {title} {\bibinfo
  {title} {Birth, life, and death of a dipolar supersolid},\ }\href
  {https://doi.org/10.1103/PhysRevLett.126.233401} {\bibfield  {journal}
  {\bibinfo  {journal} {Phys. Rev. Lett.}\ }\textbf {\bibinfo {volume} {126}},\
  \bibinfo {pages} {233401} (\bibinfo {year} {2021})}\BibitemShut {NoStop}%
\bibitem [{\citenamefont {S\'anchez-Baena}\ \emph {et~al.}(2023)\citenamefont
  {S\'anchez-Baena}, \citenamefont {Politi}, \citenamefont {Maucher},
  \citenamefont {Ferlaino},\ and\ \citenamefont {Pohl}}]{SanchezBaena2023}%
  \BibitemOpen
  \bibfield  {author} {\bibinfo {author} {\bibfnamefont {J.}~\bibnamefont
  {S\'anchez-Baena}}, \bibinfo {author} {\bibfnamefont {C.}~\bibnamefont
  {Politi}}, \bibinfo {author} {\bibfnamefont {F.}~\bibnamefont {Maucher}},
  \bibinfo {author} {\bibfnamefont {F.}~\bibnamefont {Ferlaino}},\ and\
  \bibinfo {author} {\bibfnamefont {T.}~\bibnamefont {Pohl}},\ }\bibfield
  {title} {\bibinfo {title} {Heating a dipolar quantum fluid into a solid},\
  }\href {https://doi.org/10.1038/s41467-023-37207-3} {\bibfield  {journal}
  {\bibinfo  {journal} {Nat. Commun.}\ }\textbf {\bibinfo {volume} {14}},\
  \bibinfo {pages} {1868} (\bibinfo {year} {2023})}\BibitemShut {NoStop}%
\bibitem [{\citenamefont {S\'anchez-Baena}\ \emph {et~al.}(2024)\citenamefont
  {S\'anchez-Baena}, \citenamefont {Pohl},\ and\ \citenamefont
  {Maucher}}]{SanchezBaena2024}%
  \BibitemOpen
  \bibfield  {author} {\bibinfo {author} {\bibfnamefont {J.}~\bibnamefont
  {S\'anchez-Baena}}, \bibinfo {author} {\bibfnamefont {T.}~\bibnamefont
  {Pohl}},\ and\ \bibinfo {author} {\bibfnamefont {F.}~\bibnamefont
  {Maucher}},\ }\bibfield  {title} {\bibinfo {title} {Superfluid-supersolid
  phase transition of elongated dipolar bose-einstein condensates at finite
  temperatures},\ }\href {https://doi.org/10.1103/PhysRevResearch.6.023183}
  {\bibfield  {journal} {\bibinfo  {journal} {Phys. Rev. Res.}\ }\textbf
  {\bibinfo {volume} {6}},\ \bibinfo {pages} {023183} (\bibinfo {year}
  {2024})}\BibitemShut {NoStop}%
\bibitem [{\citenamefont {He}\ \emph {et~al.}(2025)\citenamefont {He},
  \citenamefont {S\'anchez-Baena}, \citenamefont {Maucher},\ and\ \citenamefont
  {Zhang}}]{He2025}%
  \BibitemOpen
  \bibfield  {author} {\bibinfo {author} {\bibfnamefont {L.-J.}\ \bibnamefont
  {He}}, \bibinfo {author} {\bibfnamefont {J.}~\bibnamefont {S\'anchez-Baena}},
  \bibinfo {author} {\bibfnamefont {F.}~\bibnamefont {Maucher}},\ and\ \bibinfo
  {author} {\bibfnamefont {Y.-C.}\ \bibnamefont {Zhang}},\ }\bibfield  {title}
  {\bibinfo {title} {Accessing elusive two-dimensional phases of dipolar
  bose-einstein condensates by finite temperature},\ }\href
  {https://doi.org/10.1103/PhysRevResearch.7.023019} {\bibfield  {journal}
  {\bibinfo  {journal} {Phys. Rev. Res.}\ }\textbf {\bibinfo {volume} {7}},\
  \bibinfo {pages} {023019} (\bibinfo {year} {2025})}\BibitemShut {NoStop}%
\bibitem [{\citenamefont {Bomb{\'i}n}\ \emph {et~al.}()\citenamefont
  {Bomb{\'i}n}, \citenamefont {Boronat}, \citenamefont {Mazzanti},\ and\
  \citenamefont {S\'anchez-Baena}}]{Bombin2025}%
  \BibitemOpen
  \bibfield  {author} {\bibinfo {author} {\bibfnamefont {R.}~\bibnamefont
  {Bomb{\'i}n}}, \bibinfo {author} {\bibfnamefont {J.}~\bibnamefont {Boronat}},
  \bibinfo {author} {\bibfnamefont {F.}~\bibnamefont {Mazzanti}},\ and\
  \bibinfo {author} {\bibfnamefont {J.}~\bibnamefont {S\'anchez-Baena}},\
  }\bibfield  {title} {\bibinfo {title} {Creating and melting a supersolid by
  heating a quantum dipolar system},\ }\href {https://arxiv.org/abs/2506.03071}
  {\bibfield  {journal} {\bibinfo  {journal} {arXiv:}\ }\textbf {\bibinfo
  {volume} {2506.03071}}}\BibitemShut {NoStop}%
\bibitem [{\citenamefont {Lee}\ and\ \citenamefont {Yang}(1957)}]{Lee1051957}%
  \BibitemOpen
  \bibfield  {author} {\bibinfo {author} {\bibfnamefont {T.~D.}\ \bibnamefont
  {Lee}}\ and\ \bibinfo {author} {\bibfnamefont {C.~N.}\ \bibnamefont {Yang}},\
  }\bibfield  {title} {\bibinfo {title} {Many-body problem in quantum mechanics
  and quantum statistical mechanics},\ }\href
  {https://doi.org/10.1103/PhysRev.105.1119} {\bibfield  {journal} {\bibinfo
  {journal} {Phys. Rev.}\ }\textbf {\bibinfo {volume} {105}},\ \bibinfo {pages}
  {1119} (\bibinfo {year} {1957})}\BibitemShut {NoStop}%
\bibitem [{\citenamefont {Lee}\ \emph {et~al.}(1957)\citenamefont {Lee},
  \citenamefont {Huang},\ and\ \citenamefont {Yang}}]{Lee1061957}%
  \BibitemOpen
  \bibfield  {author} {\bibinfo {author} {\bibfnamefont {T.~D.}\ \bibnamefont
  {Lee}}, \bibinfo {author} {\bibfnamefont {K.}~\bibnamefont {Huang}},\ and\
  \bibinfo {author} {\bibfnamefont {C.~N.}\ \bibnamefont {Yang}},\ }\bibfield
  {title} {\bibinfo {title} {Eigenvalues and eigenfunctions of a bose system of
  hard spheres and its low-temperature properties},\ }\href
  {https://doi.org/10.1103/PhysRev.106.1135} {\bibfield  {journal} {\bibinfo
  {journal} {Phys. Rev.}\ }\textbf {\bibinfo {volume} {106}},\ \bibinfo {pages}
  {1135} (\bibinfo {year} {1957})}\BibitemShut {NoStop}%
\bibitem [{\citenamefont {Saito}(2016)}]{Saito2016}%
  \BibitemOpen
  \bibfield  {author} {\bibinfo {author} {\bibfnamefont {H.}~\bibnamefont
  {Saito}},\ }\bibfield  {title} {\bibinfo {title} {Path-integral monte carlo
  study on a droplet of a dipolar bose–einstein condensate stabilized by
  quantum fluctuation},\ }\href {https://doi.org/10.7566/JPSJ.85.053001}
  {\bibfield  {journal} {\bibinfo  {journal} {J. Phys. Soc. Jpn.}\ }\textbf
  {\bibinfo {volume} {85}},\ \bibinfo {pages} {053001} (\bibinfo {year}
  {2016})}\BibitemShut {NoStop}%
\bibitem [{\citenamefont {Schmidt}\ \emph {et~al.}(2021)\citenamefont
  {Schmidt}, \citenamefont {Hertkorn}, \citenamefont {Guo}, \citenamefont
  {B\"ottcher}, \citenamefont {Schmidt}, \citenamefont {Ng}, \citenamefont
  {Graham}, \citenamefont {Langen}, \citenamefont {Zwierlein},\ and\
  \citenamefont {Pfau}}]{Schmidt2021}%
  \BibitemOpen
  \bibfield  {author} {\bibinfo {author} {\bibfnamefont {J.-N.}\ \bibnamefont
  {Schmidt}}, \bibinfo {author} {\bibfnamefont {J.}~\bibnamefont {Hertkorn}},
  \bibinfo {author} {\bibfnamefont {M.}~\bibnamefont {Guo}}, \bibinfo {author}
  {\bibfnamefont {F.}~\bibnamefont {B\"ottcher}}, \bibinfo {author}
  {\bibfnamefont {M.}~\bibnamefont {Schmidt}}, \bibinfo {author} {\bibfnamefont
  {K.~S.~H.}\ \bibnamefont {Ng}}, \bibinfo {author} {\bibfnamefont {S.~D.}\
  \bibnamefont {Graham}}, \bibinfo {author} {\bibfnamefont {T.}~\bibnamefont
  {Langen}}, \bibinfo {author} {\bibfnamefont {M.}~\bibnamefont {Zwierlein}},\
  and\ \bibinfo {author} {\bibfnamefont {T.}~\bibnamefont {Pfau}},\ }\bibfield
  {title} {\bibinfo {title} {Roton excitations in an oblate dipolar quantum
  gas},\ }\href {https://doi.org/10.1103/PhysRevLett.126.193002} {\bibfield
  {journal} {\bibinfo  {journal} {Phys. Rev. Lett.}\ }\textbf {\bibinfo
  {volume} {126}},\ \bibinfo {pages} {193002} (\bibinfo {year}
  {2021})}\BibitemShut {NoStop}%
\bibitem [{\citenamefont {Kawaguchi}\ and\ \citenamefont
  {Ueda}(2012)}]{Kawaguchi2012}%
  \BibitemOpen
  \bibfield  {author} {\bibinfo {author} {\bibfnamefont {Y.}~\bibnamefont
  {Kawaguchi}}\ and\ \bibinfo {author} {\bibfnamefont {M.}~\bibnamefont
  {Ueda}},\ }\bibfield  {title} {\bibinfo {title} {Spinor bose–einstein
  condensates},\ }\href
  {https://doi.org/https://doi.org/10.1016/j.physrep.2012.07.005} {\bibfield
  {journal} {\bibinfo  {journal} {Phys. Rep.}\ }\textbf {\bibinfo {volume}
  {520}},\ \bibinfo {pages} {253} (\bibinfo {year} {2012})},\ \bibinfo {note}
  {spinor Bose--Einstein condensates}\BibitemShut {NoStop}%
\bibitem [{\citenamefont {Ronen}\ \emph {et~al.}(2006)\citenamefont {Ronen},
  \citenamefont {Bortolotti},\ and\ \citenamefont {Bohn}}]{Ronen2006}%
  \BibitemOpen
  \bibfield  {author} {\bibinfo {author} {\bibfnamefont {S.}~\bibnamefont
  {Ronen}}, \bibinfo {author} {\bibfnamefont {D.~C.~E.}\ \bibnamefont
  {Bortolotti}},\ and\ \bibinfo {author} {\bibfnamefont {J.~L.}\ \bibnamefont
  {Bohn}},\ }\bibfield  {title} {\bibinfo {title} {Bogoliubov modes of a
  dipolar condensate in a cylindrical trap},\ }\href
  {https://doi.org/10.1103/PhysRevA.74.013623} {\bibfield  {journal} {\bibinfo
  {journal} {Phys. Rev. A}\ }\textbf {\bibinfo {volume} {74}},\ \bibinfo
  {pages} {013623} (\bibinfo {year} {2006})}\BibitemShut {NoStop}%
\bibitem [{\citenamefont {Eberlein}\ \emph {et~al.}(2005)\citenamefont
  {Eberlein}, \citenamefont {Giovanazzi},\ and\ \citenamefont
  {O'Dell}}]{Eberlein2005}%
  \BibitemOpen
  \bibfield  {author} {\bibinfo {author} {\bibfnamefont {C.}~\bibnamefont
  {Eberlein}}, \bibinfo {author} {\bibfnamefont {S.}~\bibnamefont
  {Giovanazzi}},\ and\ \bibinfo {author} {\bibfnamefont {D.~H.~J.}\
  \bibnamefont {O'Dell}},\ }\bibfield  {title} {\bibinfo {title} {Exact
  solution of the thomas-fermi equation for a trapped bose-einstein condensate
  with dipole-dipole interactions},\ }\href
  {https://doi.org/10.1103/PhysRevA.71.033618} {\bibfield  {journal} {\bibinfo
  {journal} {Phys. Rev. A}\ }\textbf {\bibinfo {volume} {71}},\ \bibinfo
  {pages} {033618} (\bibinfo {year} {2005})}\BibitemShut {NoStop}%
\bibitem [{\citenamefont {Ronen}\ \emph {et~al.}(2007)\citenamefont {Ronen},
  \citenamefont {Bortolotti},\ and\ \citenamefont {Bohn}}]{RonenL2007}%
  \BibitemOpen
  \bibfield  {author} {\bibinfo {author} {\bibfnamefont {S.}~\bibnamefont
  {Ronen}}, \bibinfo {author} {\bibfnamefont {D.~C.~E.}\ \bibnamefont
  {Bortolotti}},\ and\ \bibinfo {author} {\bibfnamefont {J.~L.}\ \bibnamefont
  {Bohn}},\ }\bibfield  {title} {\bibinfo {title} {Radial and angular rotons in
  trapped dipolar gases},\ }\href
  {https://doi.org/10.1103/PhysRevLett.98.030406} {\bibfield  {journal}
  {\bibinfo  {journal} {Phys. Rev. Lett.}\ }\textbf {\bibinfo {volume} {98}},\
  \bibinfo {pages} {030406} (\bibinfo {year} {2007})}\BibitemShut {NoStop}%
\bibitem [{\citenamefont {Dutta}\ and\ \citenamefont
  {Meystre}(2007)}]{Dutta2007}%
  \BibitemOpen
  \bibfield  {author} {\bibinfo {author} {\bibfnamefont {O.}~\bibnamefont
  {Dutta}}\ and\ \bibinfo {author} {\bibfnamefont {P.}~\bibnamefont
  {Meystre}},\ }\bibfield  {title} {\bibinfo {title} {Ground-state structure
  and stability of dipolar condensates in anisotropic traps},\ }\href
  {https://doi.org/10.1103/PhysRevA.75.053604} {\bibfield  {journal} {\bibinfo
  {journal} {Phys. Rev. A}\ }\textbf {\bibinfo {volume} {75}},\ \bibinfo
  {pages} {053604} (\bibinfo {year} {2007})}\BibitemShut {NoStop}%
\bibitem [{\citenamefont {Wilson}\ \emph {et~al.}(2008)\citenamefont {Wilson},
  \citenamefont {Ronen}, \citenamefont {Bohn},\ and\ \citenamefont
  {Pu}}]{Wilson2008}%
  \BibitemOpen
  \bibfield  {author} {\bibinfo {author} {\bibfnamefont {R.~M.}\ \bibnamefont
  {Wilson}}, \bibinfo {author} {\bibfnamefont {S.}~\bibnamefont {Ronen}},
  \bibinfo {author} {\bibfnamefont {J.~L.}\ \bibnamefont {Bohn}},\ and\
  \bibinfo {author} {\bibfnamefont {H.}~\bibnamefont {Pu}},\ }\bibfield
  {title} {\bibinfo {title} {Manifestations of the roton mode in dipolar
  bose-einstein condensates},\ }\href
  {https://doi.org/10.1103/PhysRevLett.100.245302} {\bibfield  {journal}
  {\bibinfo  {journal} {Phys. Rev. Lett.}\ }\textbf {\bibinfo {volume} {100}},\
  \bibinfo {pages} {245302} (\bibinfo {year} {2008})}\BibitemShut {NoStop}%
\bibitem [{\citenamefont {Leggett}(1970)}]{Leggett1970}%
  \BibitemOpen
  \bibfield  {author} {\bibinfo {author} {\bibfnamefont {A.~J.}\ \bibnamefont
  {Leggett}},\ }\bibfield  {title} {\bibinfo {title} {Can a solid be
  "superfluid"?},\ }\href {https://doi.org/10.1103/PhysRevLett.25.1543}
  {\bibfield  {journal} {\bibinfo  {journal} {Phys. Rev. Lett.}\ }\textbf
  {\bibinfo {volume} {25}},\ \bibinfo {pages} {1543} (\bibinfo {year}
  {1970})}\BibitemShut {NoStop}%
\bibitem [{\citenamefont {Leggett}(1998)}]{Leggett1998}%
  \BibitemOpen
  \bibfield  {author} {\bibinfo {author} {\bibfnamefont {A.~J.}\ \bibnamefont
  {Leggett}},\ }\bibfield  {title} {\bibinfo {title} {On the superfluid
  fraction of an arbitrary many-body system at t=0},\ }\href
  {https://doi.org/10.1023/B:JOSS.0000033170.38619.6c} {\bibfield  {journal}
  {\bibinfo  {journal} {Journal of Statistical Physics.}\ }\textbf {\bibinfo
  {volume} {93}},\ \bibinfo {pages} {927–941} (\bibinfo {year}
  {1998})}\BibitemShut {NoStop}%
\end{thebibliography}%
	
\end{document}